\documentclass{aa}
\usepackage{graphicx}
\begin{document}
\title{Pre-main sequence spectroscopic binaries
suitable for VLTI observations
\thanks{based on observations obtained at the European Southern Observatory
at La Silla, Chile in program 62.I-0418(A); 63.I-0096(A);
64.I-0294(A); 65.I-0012(A); 67.C-0155(A); 68.C-0292(A);
68.C-0561(A); 69.C-0207(A); 70.C-0163(A); 073.C-0355(A); 
074.A-9018(A); 075.C-0399(A-F)}}

\authorrunning{Guenther et al.}
\titlerunning{PMS spectroscopic binaries}


   \author{E.W. Guenther\inst{1}
          \and
          M Esposito\inst{1,2}
          \and
          R. Mundt\inst{3}
          \and
          E. Covino\inst{4}
           \and
          J.M. Alcal\'a\inst{4}
           \and
          F. Cusano\inst{1}
           \and
          B. Stecklum\inst{1}
             }

\offprints{Eike Guenther, \email{guenther@tls-tautenburg.de}}

     \institute{Th\"uringer Landessternwarte Tautenburg,
              Sternwarte 5, D-07778 Tautenburg, Germany
               \and
              Dipartimento di Fisica ``E.R. Caianiello'', 
              Universit\`a  di Salerno, via S. Allende, 
                    84081 Baronissi (Salerno), 
              Italy
          \and
               Max-Planck-Institut f\"ur Astronomie,
               K\"onigstuhl 17,
              D-69117 Heidelberg,
              Germany
          \and
              INAF-Osservatorio Astronomico di Capodimonte
              via Moiariello 16,
              I-80131 Napoli,
              Italy
              }

   \date{Received 20.05.2006; accepted 6.2.2007}

\abstract{A severe problem of the research in star-formation is that the
masses of young stars are almost always estimated only from evolutionary
tracks. Since the tracks published by different groups differ, it is
often only possible to give a rough estimate of the masses of young
stars.  It is thus crucial to test and calibrate the tracks. Up to now,
only a few tests of the tracks could be carried out. However, with the
VLTI it is now possible to set constrains on the tracks by determining
the masses of many young binary stars precisely.}  {In order to use the
VLTI efficiently, a first step is to find suitable targets, which is the
purpose of this work. Given the distance of nearby star-forming regions,
suitable VLTI targets are binaries with orbital periods between at least
50 days, and few years.  Although a number of surveys for detecting
spectroscopic binaries have been carried out,  most of the binaries
found so far have periods which are too short.}  {We thus surveyed the
Chamaeleon, Corona Australis, Lupus, Sco-Cen, $\rho$ Ophiuci
star-forming regions in order to search for spectroscopic binaries with
periods longer than 50 days, which are suitable for the VLTI
observations.}  {As a result of the 8 years campaign we discovered 8
binaries with orbital periods longer than 50 days. Amongst the newly
discovered long period binaries is CS Cha, which is one of the few
classical T Tauri stars with a circumbinary disk. The survey is
limited to objects with masses higher than 0.1 to 0.2 $M_\odot$ for
periods between 1 and 8 years.}  {We find that the frequency of binaries
with orbital periods $\leq 3000\,days$ is of $20\pm5\%$. The frequency
of long and short period pre-main sequence spectroscopic binaries is
about the same as for stars in the solar neighbourhood. In total 14
young binaries are now known which are suitable for mass determination
with the VLTI.}

\maketitle


\section{Introduction}

The most fundamental parameter of a star is its mass, which determines
almost everything about its birth, life, and death. The masses of low
mass pre-main sequence (pms) stars are often derived by comparing the
location of the star in the Hertzsprung-Russell diagram with
theoretically calculated evolutionary models.  Unfortunately, the
evolutionary models published by various authors differ considerably due
to the differences in the input physics, such as the treatment of
convection, magnetic fields, chemical abundance of the star etc.
(Palla \& Stahler, \cite{palla92}; D'Antona \& Mazzitelli
\cite{dantona94}; Swenson et al. \cite{swenson94}; Burrows et
al. \cite{burrows97}; Forestini \cite{forestini94}; Siess et
al. \cite{siess97}; Baraffe et al. \cite{baraffe98}; Palla \& Stahler
\cite{palla99}; Tout et al. \cite{tout99}; Chabrier et
al. \cite{chabrier00}, D'Antona et al. \cite{dantona00}; Wuchterl \&
Tscharnuter, \cite{wuchterl03}; Montalb\'an et al. \cite{montalban04}).
Testing and calibrating evolutionary models of pms-stars are thus of key
importance for the understanding of young stars, the determination of
the initial mass function, and studies of the galactic star-formation
history in general. The lack of knowledge, which of the models to
choose, is thus a severe problem for the whole field of
star-formation. Classical T\,Tauri stars (henceforth called CTTSs) are
young, low mass, optically visible pms-emission line stars with an
accretion disk. Weak-line T Tauri stars (WTTSs) are similar to the
classical ones but have much smaller accretion rates, and less massive
disks, if any.

An ideal test of the tracks would be to compare the true masses of CTTSs
and WTTSs with the masses derived from the models. This is, for example,
possible for eclipsing binary stars. Unfortunately, by now there are
only three eclipsing young binaries known after we eliminated
\object{RXJ 1608.6-3922} by showing that it is not an eclipsing binary
but a spotted single star (Joergens et al. \cite{joergens01}). Casey et
al. (\cite{casey98}) derived the masses of the eclipsing binary
\object{TY CrA} to $3.16\pm0.02\,{\rm M}_\odot$ and $1.64\pm0.01\,{\rm
M}_\odot$ and compared these values with three sets of models.
Evolutionary models do not differ too much in the mass-range between 1.5
and 3.0 ${\rm M}_\odot$, and thus the authors find that all three sets
of models are reasonably consistent with the observations.  Since many
pms-stars have masses smaller than 1.5 ${\rm M}_\odot$ it is important
to have direct mass determinations also in the low-mass domain.
It is thus better to focus on stars with masses lower than 1.5 ${\rm
M}_\odot$.  Covino et al. (\cite{covino04}) analysed the eclipsing
binary \object{RXJ 0529.4+0041} which consists of a $1.25\,{\rm
M}_\odot$ primary star and a $0.91\,{\rm M}_\odot$ companion star and
compared these values with three sets of evolutionary models. They find
that the models published by Baraffe et al. (\cite{baraffe98}) and
Swenson et al.  (\cite{swenson94}) are in reasonable agreement with the
observations but the ones published by D'Antona \& Mazzitelli
(\cite{dantona94}) are not.  Stassun et al. (\cite{stassun04}) analysed
the eclipsing binary V1174 Ori which consists of a $1.01\pm0.02\,{\rm
M}_\odot$ primary and a $0.731\pm0.008\,{\rm M}_\odot$ secondary.  They
find that models by Montalb\'an et al. (\cite{montalban04}) seem to
agree with the observations but those published by Siess et
al. (\cite{siess97}) are in conflict with them.

By measuring the orbital motion of the molecular gas in the disk, Simon
et al. (\cite{simon00}) concluded that models of the pms-evolution with
lower $T_{\rm eff}$ values are in better agreement with the observations
(see, e.g., Baraffe et al. \cite{baraffe98} and Palla \& Stahler
\cite{palla99}).  Another possibility to measure masses directly is to
combine astrometric data obtained with the HST Fine Guidance Sensors
with radial velocity (RV) measurements. In this way Steffen et
al. (\cite{steffen01}) derived masses of $1.5\pm0.2\,{\rm M}_\odot$ and
$0.81\pm0.09\,{\rm M}_\odot$ for the binary system \object{NTT
045251+3016}.  They find that the Baraffe et al.  (\cite{baraffe98})
models with a mixing-length parameter of $\alpha=1.0$ are closest to the
measured primary mass. The models published by D'Antona \& Mazzitelli
(\cite{dantona94}) are clearly inconsistent with the observations. The
first determination of the masses of the components of a young binary
star using an interferometer was recently carried out by Boden et
al.(\cite{boden05}).  They used the Keck interferometer and determined
the masses of the two components of \object{HD\,98800\,B} to
$0.70\pm0.06$ $M_\odot$ and $0.58\pm0.05$ $M_\odot$. With an age of 8 to
20 Myr it is however much older than a typical T\,Tauri star.
Nevertheless, the object is still young enough to be used for testing
the tracks. However, a reasonable agreement with the models published by
Baraffe (\cite{baraffe98}) could only be achieved by assuming a rather
low metallicity of the star.

\begin{figure}[h]
\includegraphics[width=0.35\textwidth, angle=270]{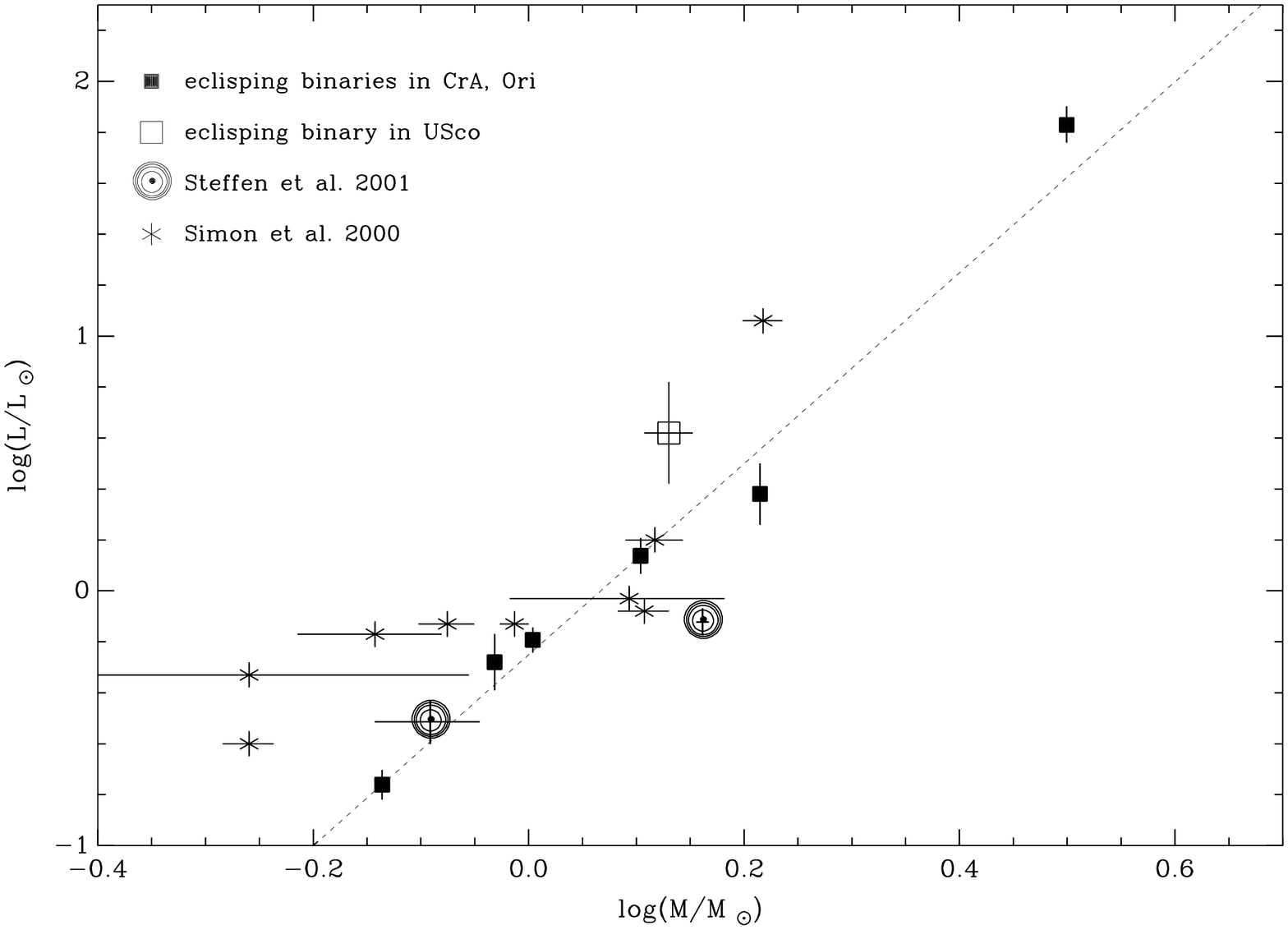}
\caption{The figure summarises the current knowledge of empirical mass
determination of young stars.  The masses of the eclipsing objects in
Orion and CrA were taken from Covino et al. (\cite{covino04}), Stassun
et al. (\cite{stassun04}), and Casey et al. (\cite{casey98}).  Alencar
et al. (\cite{alencar03}) derived the masses of the binary in upper
Sco. Using the HST Fine Guidance Sensor Steffen et
al. (\cite{steffen01}) determined the masses of NTT 045251+3016 in
CrA. Also shown are the masses derived by Simon et al. (\cite{simon00}) by
measuring the orbital motion of gas in the disk.}
\label{luminosity_mass}
\end{figure}

While all these data allow a very first test of evolutionary models,
more mass determinations are definitely needed, especially of stars of
lower masses.  Shown in Fig.\,\ref{luminosity_mass} is $log(L/L_\odot)$
versus $log(M/M_\odot)$ for all young stars of which the masses have so
far been derived. However, we should keep in mind that the position of
young stars in the Hertzsprung-Russell diagram depends on the history of
the accretion rate.  Thus, short-period, eclipsing binaries might not be
ideal for testing the tracks, because the components might have had
recent mass-exchange.  An example where this could have been the case is
\object{RXJ1603.8-3938}. In this system both components have almost
identical masses but one is a factor two brighter than the other
(Guenther et al. \cite{guenther01}).

In order to determine the masses of many more pms-stars over a
large mass range we have carried out a survey for long-period binary
stars.  For single-line spectroscopic binaries (SB1) only the
mass-function $f(m)={{m_2^3\sin^3{i}}\over{(m_1+m_2)^2}}$ can be derived
using the RV-method. If $f(m)$ is combined with the relative astrometric
orbit of the two stars (position angle and relative distance of the
stars) and if also its parallax is known, the masses of the two stars
can be determined. Using this method, we have derived the masses of the
two components of the SB1-system \object{EK Dra} (K\"onig et al.
\cite{koenig05}). This system has, however, an age of about 125 Myr and
thus, it is not really young.  For double-line spectroscopic binaries (SB2),
it is possible to measure $m_1\,sin^3\,i$ and $m_2\,sin^3\,i$. The
individual masses of the stars can then be derived if the RV-data is
combined with the relative astrometric orbit, or if the system is
eclipsing.

Unfortunately, the distance to the nearest star-forming regions is so
large, that binaries that could be resolved with AO-systems on 8 m-class
telescopes have orbital periods of many decades, or even centuries.
However, the big leap forward in this work is the Very Large Telescope
Interferometer, which is now available. The VLTI instrument AMBER
(Astronomical Multi BEam combineR) offers the unique opportunity to
spatially resolve binaries down to separations of only 4 mas in the
K-band. Thus, with AMBER it is possible to resolve binaries with
separations down to 0.6~AU (corresponding to an orbital period of
$\sim$ 120~d for systems with $m_1=m_2=M_\odot$) for distances $\leq$
150~pc.  Because AMBER works in the near infrared, it is possible to
detect a young $0.1\,{\rm M}_\odot$ star next to a young $1.0\,{\rm
M}_\odot$ star due to the achievable brightness contrast ratio.

In order to carry out such measurements with AMBER a sufficient number
of pms-binaries with suitable separations (i.e. sufficiently long
periods) have to be identified first. The aim of this work thus is the
identification of pms-binaries suitable for the VLTI observations.  For
the survey we selected known pms-stars in southern star-forming regions
which have a distance $\leq$ 150 pc. Stars fainter than 10 mag in
the K-band, visual pairs with a separation of $\leq$ 10 arcsec and stars
with a $v\,sin\,i\geq 50$ $km\,s^{-1}$, are excluded.

Before we started our survey in 1998, only three pms-spectroscopic
binaries located in nearby star-forming regions and with orbital periods
longer than 50 days were known (Mathieu \cite{mathieu94}).  Four
additional spectroscopic binary {\em candidates} in the
Ophiuchus-Scorpius (Oph), Chamaeleon, Lupus, and Corona Australis star
forming regions were identified by Melo et al. (\cite{melo03}).  Thus,
while quite a number of authors have already surveyed the nearby
star-forming regions for spectroscopic binaries, most of the objects
found are binaries of short period and unsuitable for VLTI observations.
The main intention of our survey is a search for long-period binaries,
suitable for VLTI-observations, not a comprehensive survey of
spectroscopic binaries. In this paper we to present the outcome of this
extensive survey. The short-period systems, which were also discovered
in the course of the survey will be subject of a forthcoming paper.  In
section 2, we describe the observations, in section 3, we give a full
list of all long period pms-binaries suitable for the VLTI observations
together with the RV-values and the orbits of the newly discovered
binaries. In section 4, we discuss the spectroscopic binary candidates.
In the last section we discuss the results obtained.

\section{The sample}

For our survey we selected 122 late-type pms-stars in the Chamaeleon
(Cha), Corona Australis (CrA), $\rho$ Ophiuchi (Oph), Lupus (Lup), and
Scorpius Centaurus (SC) star-forming regions. We did not include the 12
SBs known at the star of the survey (Mathieu et al. \cite{mathieu94};
Melo \cite{melo03}; Guenther et al. \cite{guenther01}), and also did in
general not include visual binaries, unless the separation was $\geq$ 10
arcsec. After the first observing run, we removed another 14 stars from
the list, because their $v\,sin\,i$ turned out to be too high, or
because these stars are not young. Apart from the known binaries,
and stars with very large $v\,sin\,i$m, we observed essentially all stars
with spectral types between F6 and M3 in these regions.  The sample
that was finally observed comprises 108 stars, of which 26 are
CTTS, and 82 WTTS.

Determinations of the distance of the Chamaeleon association give values
of $160\pm15$~pc for ChaI, $178\pm18$~pc for ChaII (Whittet et al.
\cite{whittet97}), $171\pm20$~pc for ChaI (Wichmann et
al. \cite{wichmann98}), and $168^{+14}_{-12}$ pc (Bertout et
al. \cite{bertout99}). For the Corona Australis region Knude \& H\o g
(\cite{knude98}) derive a distance of 170 pc which is different from the
$\sim$ 130 pc found by other authors (Knacke et al. \cite{knacke73};
Marraco \& Rydgren \cite{marraco81}).  For the $\rho$ Ophiuchi region,
the distance of 160 pc derived by Knude \& H\o g (\cite{knude98}) agrees
well with the results obtained by other authors (Whittet
\cite{whittet74}; Chini \cite{chini81}).  Quite a number of authors have
determined the distance to the Lupus star-forming region: Hughes et
al. (\cite{hughes93}) find $140\pm20$~pc, Knude \& H\o g
(\cite{knude98}) 100~pc, Nakajima et al. (\cite{nakajima00}) 150 pc,
Sartori et al. (\cite{satori03}) 147 pc, Franco (\cite{franco02}) 150
pc, de Zeeuw et al. (\cite{dezeeuw99}) $142\pm2$~pc, and Teixeira et
al. (\cite{teixeira00}) 85~pc but note that 14 stars of this group have
measured parallax-distances, with an average of 138~pc.

The Upper Scorpius OB association has an age of 5-6 Myr (Preibisch \&
Zinnecker \cite{preibisch99}) and thus is slightly older than the other
regions that all have ages of 1-3 Myr. The average distance of the Upper
Scorpius OB association is $145\pm2$ pc (de Zeeuw et
al. \cite{dezeeuw99}). In Tab.\,\ref{tab:binariesI} we also give the
parameters of the binaries found in the TW Hydra (TWA) and Tucana (Tuc)
associations. Both regions are older but also closer (45-60 pc) than the
other regions.  Since low-mass stars in these regions are still pms,
these objects are interesting for VLTI observations.  The TW Hydra
region has a age of 8-12 Myr, and the Tucana association an age of about
30 Myr (Webb et al. \cite{webb99}; Torres et al. \cite{torres00}; Torres
et al. \cite{torres03}; Weinberger et al. \cite{weinberger04}). 

\section{Observations}

All stars were observed with the ESO Echelle spectrograph FEROS
(Fiber-fed Extended Range Optical Spectrograph).  FEROS was operated up
to October 2002 (HJD 245\,2550) at the ESO 1.5-m-telescope, and was then
moved to the MPG/ESO-2.20-m-telescope.  The spectra cover the wavelength
region between about 3600 \AA \ and 9200 \AA \, with a resolving power
of $\rm \lambda / \Delta \lambda=48000$. On the 1.5-m-telescope the
entrance aperture of the fibre had a projected diameter of $2 \farcs 7$
on the sky, while at MPG/ESO 2.2-m-telescope $2 \farcs 0$. As long
as FEROS was at the ESO 1.5-m-telescope, exposure-times were set so that
a S/N-ratio of about 30 was achieved. The exposure-times were thus
between 5 and 45 minutes. Since the performance of FEROS dramatically
improved when FEROS was moved to the MPG/ESO 2.2-m-telescope, the
S/N-ratio went up to typically 50, although we shortened the exposure
times. The standard FEROS data-reduction pipeline was used for bias
subtraction, flat-fielding, scattered light removal, Echelle order
extraction, and wavelength calibration of the spectra.

When measuring the radial velocity of stars, the question arises
whether it is better to use several templates of different spectral
types, or only one template. We tried out both methods using
\object{HR3862} (G0), \object{HR6748} (G5), \object{HR 5777} (K1),
\object{HR5568} (K4), and \object{HR6056} (M0.5) as templates. The
advantage in using several templates is that the match between the
template and the star is better, but the disadvantage is that the
absolute RVs of the templates have to be known to a high accuracy, if
the absolute RVs of a stars are of interest. If the absolute RVs of the
templates are not well known to a high accuracy, the absolute values of
the RVs of the stars are lost. In our case this would be a disadvantage,
because the first step for detecting pms-spectroscopic binaries is to
determine their absolute RV. A star with a RV that differs by more than
2 $km\,s^{-1}$ from that of the star-forming region is likely to be
either not a member of that region, or an SB1.

\begin{figure}[h]
\includegraphics[width=0.35\textwidth, angle=270]{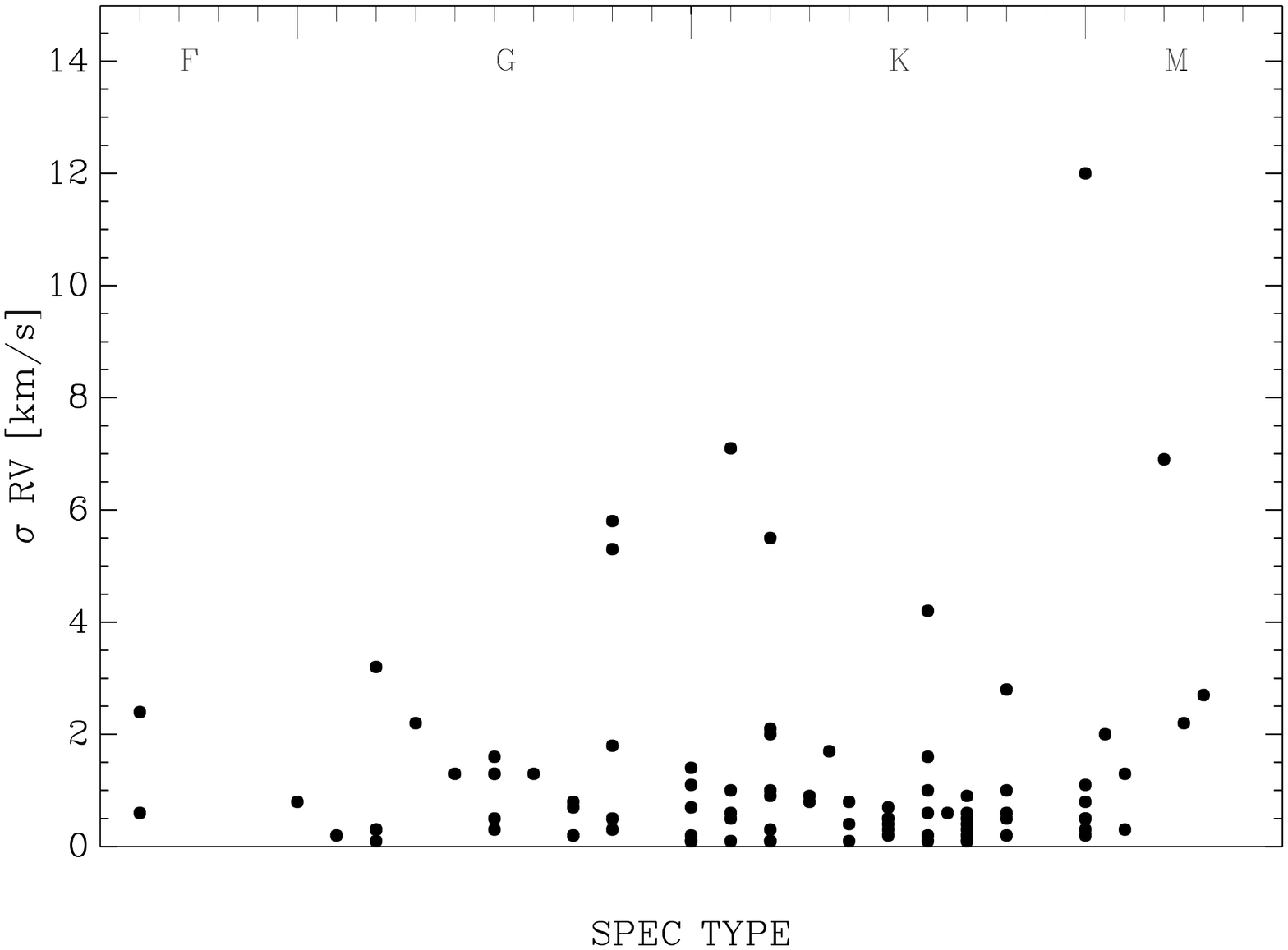}

\caption{The figure shows the scatter of the measurements of all single
stars listed in Table\,\ref{tab:singleI}, \ref{tab:singleII},
\ref{tab:singleIII} versus the spectral type. Assuming that the scatter
is the error of the measurements, we conclude that the error dominated by
the v\,sin\,i of the stars, and there is no obvious trend with the spectral
type.}
\label{SpecType_error}
\end{figure}

The other alternative is to use just one template. This approach is
used in the HARPS search for extra-solar planets of F,G, and K-stars,
where an accuracy of typically 1 $m\,s^{-1}$ is achieved (Lovis et
al. \cite{lovis05}).  A good choice in our case is the K1-star
\object{HR 5777}, which is bright ($m_v=4.6$), and its absolute RV is
well determined ($+49.12\pm0.06$ $kms^{-1}$, Murdoch et
al. \cite{murdoch93}).  If we use just one template, the question is
whether the mismatch of the spectral type between the template and the
star reduces the accuracy of the RV-measurements. If this were the case,
the accuracy of the RV-measurements of the K-stars would be higher than
those of stars with other spectral types.  Fig.\,\ref{SpecType_error}
shows the variance of the RV-measurements versus the spectral type.
There is no obvious trend of the variance with the spectral type.  This
is possibly because we have only very few M-stars in our survey, were
the mismatch with the template really matters. We thus use only
\object{HR 5777} as template. In this way, it is possible to compare the
RVs of different stars, and it is possible to merge our data with data
taken with other instruments.

We split up the spectrum in six wavelength regions which are practically
free of stellar emission lines, and virtually free of telluric
lines. The wavelength regions are: 4000 to 4850 \AA, 4900 to 5850 \AA,
5900 to 6500 \AA, 6600 to 6860 \AA, 7400 to 7500 \AA, and 7700 to 8100
\AA. For each of the wavelength-band we obtained the RV separately, and
then averaged these 6 values.  The errors of the RV-measurements of
the T Tauri stars are determined from the variance of the RV-values of
the individual spectral regions.  In many stars, we could use only the
first three regions, because the S/N ratio of the other regions was too
low. In such cases we always used the same regions for the same star. In
order to account for any instrumental shift, we measured the position of
the telluric spectral lines in the 6860 to 6930 \AA \, region by
cross-correlating this part of the stellar spectrum with one taken with
the Fourier Transform Spectrometer of the McMath-Pierce telescope at
Kitt Peak (Wallace et al. \cite{wallace98}; Brault \cite{brault78}). We
find an instrumental shift of typically 0.5 $km\,s^{-1}$ which 
differs typically by less than 0.1 $km\,s^{-1}$ from frame to frame.

\begin{figure}[h]
\includegraphics[width=0.35\textwidth, angle=270]{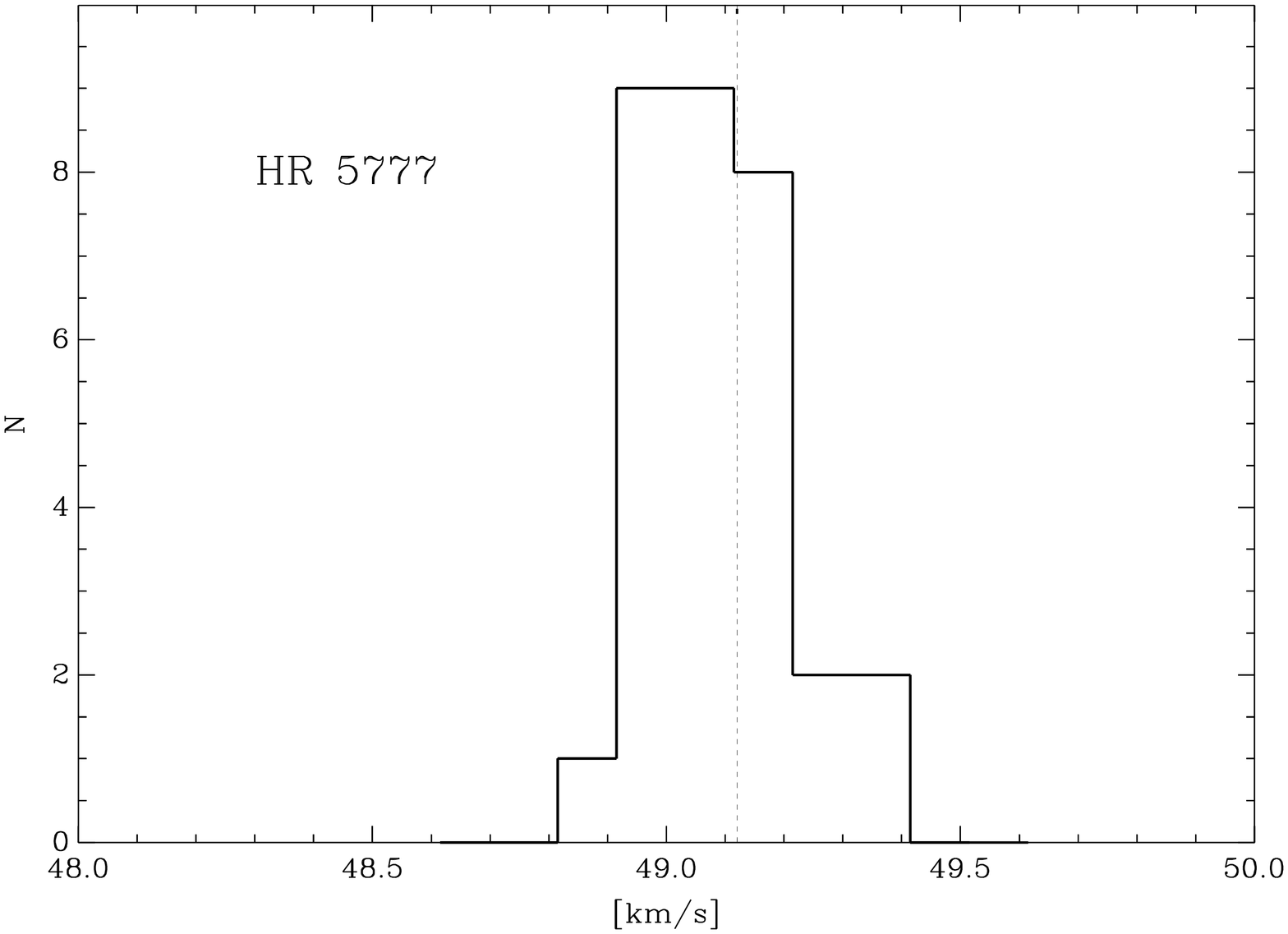}

\caption{The figure shows a histogram of the RV measurements obtained
for \object{HR 5777}. The published RV of this star is $+49.12\pm0.06$
$kms^{-1}$ (Murdoch \cite{murdoch93}) which is shown as a dashed
line. The FEROS measurements are consistent with this value.  The
FWHM of a Gaussian fitted to this the distribution is 0.30
$kms^{-1}$. The error of the measurements thus is $\pm$ 0.15
$kms^{-1}$.}
\label{HR5777}
\end{figure}

In order to investigate the accuracy of the RV-measurements obtained, we
took 31 spectra of \object{HR 5777} spread out over the whole observing
campaign.  These spectra were reduced in the same way as the spectra of
our targets. That is, we used one spectrum of \object{HR 5777} as
template (taken on the March 23, 1999), and cross-correlated this
spectrum with all other spectra taken of \object{HR 5777}. The absolute
RV of the template was determined by cross-correlating it with the
FTS-spectrum of the sun which was rebinned to the same resolution
as the FEROS spectra.  (Wallace et al. \cite{wallace98}; Brault
\cite{brault78}).  Fig.\,\ref{HR5777} shows a histogram of the
RV-measurements of \object{HR 5777}.  The RV-values determined for
\object{HR 5777} are perfectly consistent with the published RV of
$+49.12\pm0.06$ $kms^{-1}$ (Murdoch et al. \cite{murdoch93}). 
Fitting a Gaussian to this distribution, we derive a FWHM of 0.30
$kms^{-1}$. This implies that the error is $\pm0.15$ $kms^{-1}$, which
agrees well with the error derived from the variance of the RV-values
which is $\pm0.12$ $kms^{-1}$. We thus take $\pm0.15$ $kms^{-1}$ as the
intrinsic uncertainty of the measurements. The uncertainty of the
RV-measurements of the pre-main sequence stars as determined from
the variance of the RV-measurements of the individual spectral regions,
are usually larger than those of \object{HR 5777}, because of the
smaller signal-to-noise and the larger $v\,sin\,i$ of the stars.
 
\section{Results: Long-period spectroscopic binaries}

The aim of this survey is to detect young binaries suitable for
VLTI-observations, i.e. binaries with orbital periods longer than 50
days located in the nearby star-forming regions. While in our optical
observations these systems usually appear as SB1s, many of these can be
converted into SB2s, if high-resolution infrared spectra were taken
(e.g. Prato et al. \cite{prato02}).  Table\,\ref{tab:binariesI} gives a
short overview of all known long-period binaries in nearby southern
star-forming regions found during this study (\object{CS Cha},
\object{RXJ1220.6-7539}, \object{MO Lup}, \object{RXJ1534.1-3916},
\object{RXJ1559.2-3814}, \object{GSC 06209-00735}, \object{GSC
06213-00306}, \object{BS Indi}) and by other authors.  In the following,
we give details on the individual systems listed in
Table\,\ref{tab:binariesI}.  Additionally to the binaries found in this
survey, we also list in Tab.\,\ref{tab:binariesI} the long-period
spectroscopic binaries published elsewhere (\object{HIP50796}, Torres et
al. \cite{torres03}; \object{HD97131}, Torres et al. \cite{torres03};
\object{NTTS160814-1857}, Mathieu \cite{mathieu94}; \object{Haro 1-14c},
Reipurth et al. \cite{reipurth02}, Simon \& Prato \cite{simon04};
\object{NTTS162819-2423s}, Mathieu \cite{mathieu94}).  While we can not
completely rule out that the \object{VW Cha} and \object{GSC
06793-00569} are binaries, these objects are in any case unsuitable for
the VLTI observations, and thus not listed in Tab.\,\ref{tab:binariesI}.

\begin{table*}
\caption{Spectroscopic binaries with periods longer than 50 days}
\begin{tabular}{lccclcclllr}
\hline \hline
                     & this srvey/ &  EW H$\alpha$ & EW LiI & spec type & $m_K$ & RA & Dec & type & period \\
                     & region      &        & [\AA]$^1$     & [\AA]  & type & [mag] & (2000.0) & (2000.0) &  & [days] \\
\hline
HIP50796$^2$         & no / TWA  & $0.20^3$ &                 & K5/WTTS & $7.66\pm0.03$ & 10 22 18.0 & -10 32 15 & SB1 & 570 \\
CS Cha               & yes / Cha &      -40 & $0.53\pm0.01$   & K4/CTTS & $8.20\pm0.03$ & 11 02 26.3 & -77 33 36 & SB1 & $\geq$ 2482 \\
HD97131$^4$          & no / TWA  &          &                 & F2      & $7.70\pm0.02$ & 11 10 34.2 & -30 27 19 & ST3 & 134 \\
RXJ1220.6-7539       & yes / Cha & fi       & $0.21\pm0.06$   & K2/WTTS & $7.93\pm0.02$ & 12 20 34.4 & -75 39 29 & SB1 & 613 \\
MO Lup$^5$           & yes / Lup & -2.3     & $0.37\pm0.02$   & K7/WTTS & $8.64\pm0.02$ & 15 24 03.5 & -32 09 51 & ST3 & $>$ 3000 \\
RXJ1534.1-3916       & yes / Lup & abs      & $0.21\pm0.02$   & K1/WTTS & $8.55\pm0.02$ & 15 34 07.4 & -39 16 18 & SB1 & $>$ 3000 \\
RXJ1559.2-3814       & yes / Lup & -1.4     & 0.23/0.14       & WTTS    & $9.34\pm0.03$    & 15 59 16.1 & -38 14 42 & SB2 & 474 \\
GSC 06209-00735      & yes / SC  &  0.3     & $0.37\pm0.01$   & K2/WTTS & $8.43\pm0.02$ & 16 08 14.8 & -19 08 33 & SB1 & 2045 \\
NTTS160814-1857$^6$  & no / SC   & 0.7      &                 & K2/WTTS & $7.69\pm0.02$ & 16 11 09.0 & -19 04 45 & SB1 & 145 \\
GSC 06213-00306      & yes / SC  & fi       & 0.24/0.18       & WTTS    & $7.43\pm0.02$ & 16 13 18.5 & -22 12 48 & SB2 & 167 \\
Haro 1-14c$^7$       & no /  Oph &          &                 & K3/WTTS & $7.78\pm0.03$ & 16 31 04.4 & -24 04 33 & SB2 & 591 \\
NTTS162819-2423s$^6$ & no / Oph  & em       &                 & G8/WTTS & $7.44\pm0.02$ & 16 31 20.0 & -24 30 04 & SB1 & 89 \\ 
BS Indi$^8$          & yes / Tuc & abs      & $0.18\pm0.02$   & K0/WTTS & $6.57\pm0.02$ & 21 20 59.8 & -52 28 40 & SB1 & 1222 \\
\hline\hline
\end{tabular}
\label{tab:binariesI}
$^1$ em=emission, abs=absorption fi=filled in\\
$^2$ triple system Torres et al. (\cite{torres03}) \\
$^3$ Song et al. (\cite{song02}) \\
$^4$ Torres et al. (\cite{torres03}), possibly also a triple system \\
$^5$ triple system consisting of a binary with 12 days, and one of with a period $>$ 3000 days.
(Esposito et al. \cite{esposito06}).\\
$^6$ Mathieu (\cite{mathieu94}) \\
$^7$ Reipurth et al. (\cite{reipurth02}), Simon \& Prato(\cite{simon04}),
     eccentricity $0.617\pm0.008$, $f(m)=0.018\pm0.001\,M_\odot$ \\
$^8$ triple system, see Guenther et al. (\cite{guenther05}) \\ 
\end{table*}

\subsection{CS Cha}

\begin{figure}[h]
\includegraphics[width=0.55\textwidth, angle=0]{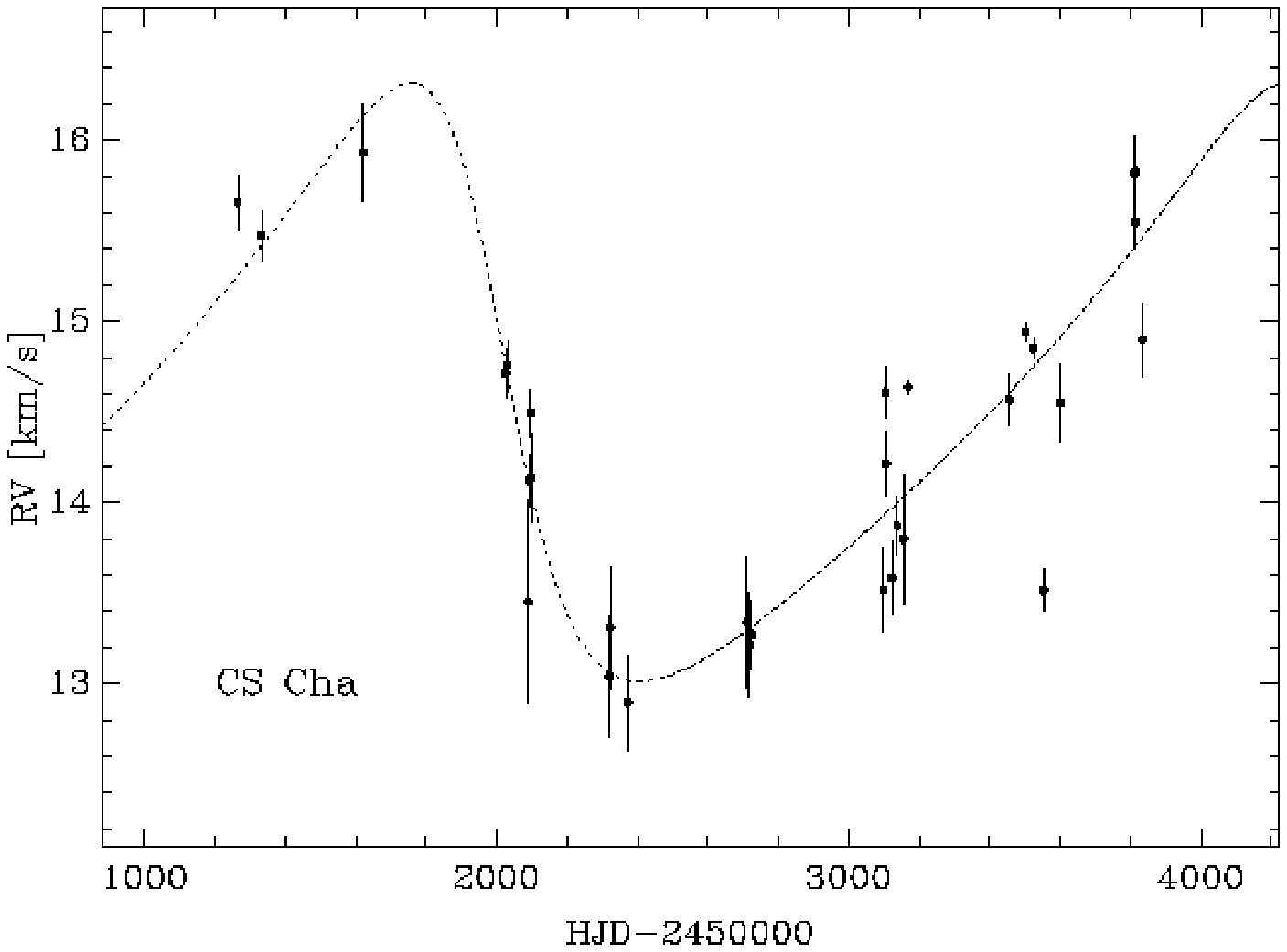}
\caption{The figure shows the RV-measurements of the CTTS \object{CS
Cha}, together with a possible orbital solution with a period of 2482
days. However, we can not rule out that the period is even longer than
that.}
\label{cs_cha}
\end{figure}

\object{CS Cha} is a CTTS of solar abundance (Padgett
\cite{padgett96}).  This object was observed by Ghez et
al. (\cite{ghez97}) by means of speckle-imaging but no companion was
found within $0 \farcs 1$.  However, Takami et al. (\cite{takami03})
already conclude from the fact that there is a large gap in the inner
disk that this object might actually be a binary. The basic properties
of the disk of this star were derived by Henning et al.
(\cite{henning93}).  We took 32 spectra of this object over a period of
2570 days Table\,\ref{RV_cs_cha}. \object{CS Cha} is an SB1-binary
system with a very long period. A possible orbit of 2482 days is shown
together with the RV-measurements in Fig.\,\ref{cs_cha}. The minimum
mass of the companion in this case would be only about 0.1
$M_\odot$. However, it is still possible that the orbital period is even
longer than 2482 days.

\begin{table}
\caption{CS Cha}
\begin{tabular}{cc}
\hline \hline
HJD & RV [$km\,s^{-1}$]\\
\hline
245\,1264.54715 & $15.7\pm0.2$ \\
245\,1332.55406 & $15.5\pm0.1$ \\
245\,1621.62514 & $15.9\pm0.3$ \\
245\,2026.54612 & $14.7\pm0.1$ \\
245\,2031.51533 & $14.8\pm0.1$ \\
245\,2089.60062 & $13.5\pm0.6$ \\
245\,2093.54486 & $14.1\pm0.1$ \\
245\,2097.50960 & $14.5\pm0.1$ \\
245\,2098.55118 & $14.1\pm0.2$ \\
245\,2319.67764 & $13.0\pm0.3$ \\
245\,2322.65188 & $13.3\pm0.3$ \\
245\,2373.56573 & $12.9\pm0.3$ \\
245\,2710.51867 & $13.3\pm0.4$ \\
245\,2717.55061 & $13.2\pm0.3$ \\
245\,2723.53930 & $13.3\pm0.2$ \\
245\,3097.51064 & $13.5\pm0.2$ \\
245\,3104.56677 & $14.6\pm0.1$ \\
245\,3105.53213 & $14.2\pm0.2$ \\
245\,3122.55011 & $13.6\pm0.2$ \\
245\,3137.56055 & $13.7\pm0.2$ \\
245\,3155.57941 & $13.7\pm0.3$ \\
245\,3455.56786 & $14.6\pm0.1$ \\
245\,3168.54063 & $14.6\pm0.1$ \\
245\,3501.61418 & $14.9\pm0.1$ \\
245\,3524.53172 & $14.9\pm0.1$ \\
245\,3552.52001 & $13.5\pm0.1$ \\
245\,3601.52246 & $14.6\pm0.2$ \\
245\,3811.53842 & $15.8\pm0.2$ \\
245\,3811.58558 & $15.8\pm0.2$ \\
245\,3811.60357 & $15.8\pm0.2$ \\
245\,3813.78161 & $15.6\pm0.2$ \\
245\,3834.09873 & $14.9\pm0.2$\\
\hline\hline
\end{tabular}
\label{RV_cs_cha}
\end{table}

\subsection{RXJ1220.6-7539}

\begin{figure}[h]
\includegraphics[width=0.48\textwidth, angle=0]{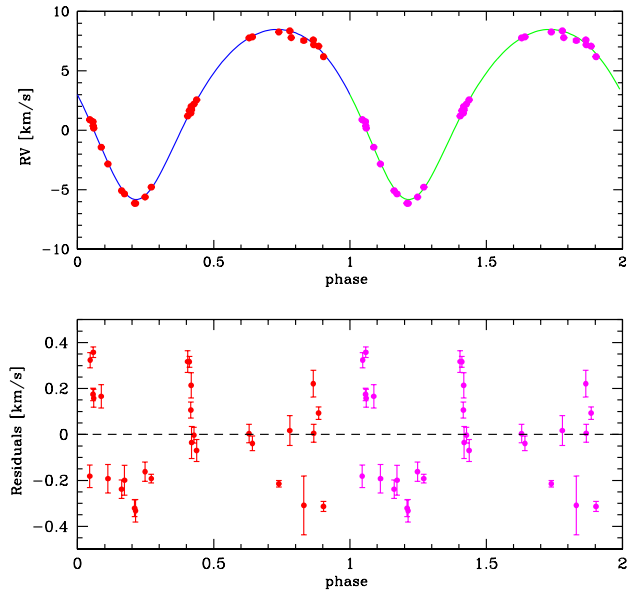}
\caption{The figure shows the phase-folded RV-measurements of
\object{RXJ1220.6-7539}. The orbital period is $612.9\pm0.4$ days.}
\label{rxj1220.6-7539}
\end{figure}

RXJ1220.6-7539 is a WTTS. In most of our spectra the $H\alpha$ line is
filled in. In some occasions we observe a small double line emission
profile with an equivalent width of only $-0.4$ \AA. At other occasions,
$H\alpha$ appears in absorption but the equivalent width then is only
0.08 \AA.  We took 33 spectra of this star over a time-span of 2587
days. The object is an SB1-binary with an orbital period of
$613.9\pm0.4$ days and a eccentricity of
$0.225\pm0.005$. Fig.\,\ref{rxj1220.6-7539} shows the phase-folded
RV-measurements. The RV-measurements are given in
Tab.\,\ref{RV_rxj1220.6-7539}, and the orbital elements in
Tab.\,\ref{orbit_rxj1220.6-7539}.

\begin{table}
\caption{RXJ1220.6-7539}
\begin{tabular}{cr}
\hline \hline
HJD & RV [$km\,s^{-1}$]\\
\hline
245\,1262.64298 & $0.7\pm0.1$ \\
245\,1333.65495 & $-5.4\pm0.1$ \\
245\,1621.67014 & $7.9\pm0.1$ \\
245\,1737.48553 & $7.5\pm0.1$ \\
245\,2089.58864 & $1.2\pm0.1$ \\
245\,2093.53472 & $1.7\pm0.1$ \\
245\,2097.55429 & $2.0\pm0.1$ \\
245\,2098.52034 & $1.7\pm0.1$ \\
245\,2319.78210 & $8.4\pm0.1$ \\
245\,2372.63728 & $7.6\pm0.1$ \\
245\,2373.61486 & $7.2\pm0.1$ \\
245\,2384.59905 & $7.1\pm0.1$ \\
245\,2395.67658 & $6.2\pm0.1$ \\
245\,2710.65545 & $1.4\pm0.1$ \\
245\,2717.69773 & $2.2\pm0.1$ \\
245\,2723.66034 & $2.6\pm0.1$ \\
245\,3096.66843 & $0.9\pm0.1$ \\
245\,3097.71156 & $0.9\pm0.1$ \\
245\,3104.70590 & $0.4\pm0.1$ \\
245\,3105.63828 & $0.2\pm0.1$ \\
245\,3122.64088 & $-1.4\pm0.1$ \\
245\,3137.72151 & $-2.8\pm0.1$ \\
245\,3168.57480 & $-5.1\pm0.1$ \\
245\,3455.66760 & $7.8\pm0.1$ \\
245\,3522.60916 & $8.3\pm0.1$ \\
245\,3550.54048 & $7.8\pm0.1$ \\
245\,3811.55386 & $-6.1\pm0.1$ \\
245\,3813.79633 & $-6.2\pm0.1$ \\
245\,3835.16435 & $-5.6\pm0.1 $ \\      
245\,3849.21861 & $-4.8\pm0.1$ \\
\hline\hline
\end{tabular}
\label{RV_rxj1220.6-7539}
\end{table}

\begin{table}
\caption{Orbital elements of RXJ1220.6-7539}
\begin{tabular}{ll}
\hline \hline
element & value \\
\hline
P           & $613.9\pm0.4$ $d$ \\
$T_0$ [HJD] & $2450123\pm3$ \\
$\gamma$       & $2.74\pm0.03$ $km\,s^{-1}$ \\
$K_{1}$     & $7.15\pm0.07$ $km\,s^{-1}$ \\
e           & $0.225\pm0.005$ \\
$\omega$    & $171.7\pm1.9^o$ \\
$a_1 sin\,i$ & $0.399\pm0.005$ AU \\
f(m) & $0.0228\pm0.0008$ $M_\odot$\\
\hline\hline
\end{tabular}
\label{orbit_rxj1220.6-7539}
\end{table}



\subsection{RXJ1534.1-3916}

\begin{figure}[h]
\includegraphics[width=0.50\textwidth, angle=0]{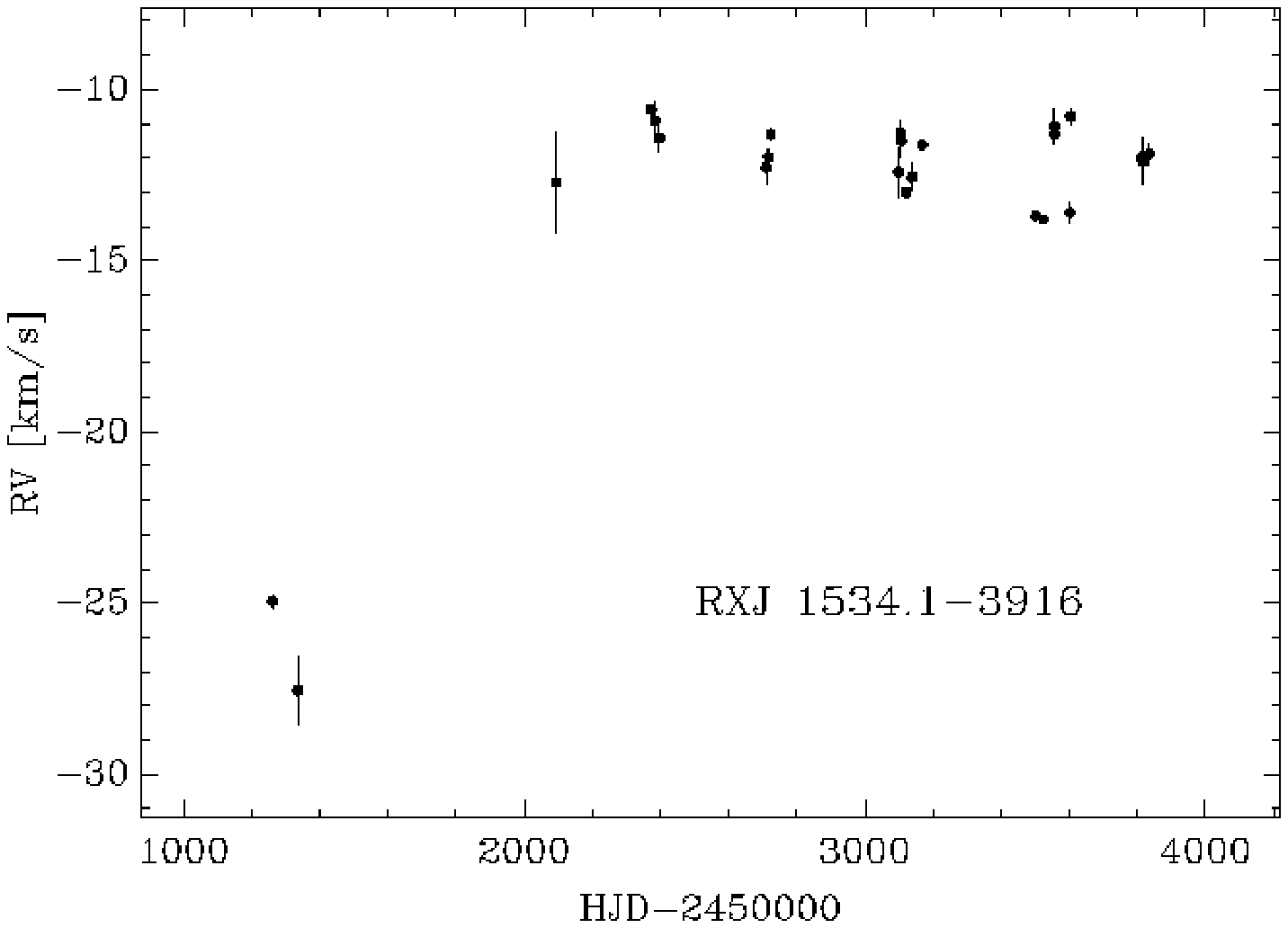}
\caption{The figure shows the RV-measurements of \object{RXJ1534.1-3916}
The orbital period is apparently longer than 7 years.}
\label{rxj1534.1-3916}
\end{figure}

\object{RXJ1534.1-3916} is a WTTS, with $H\alpha$ being in
absorption. We took 24 spectra of this star over a time-span of 2574
days. The RV-values are shown in Tab.\,\ref{RV_rxj1534.1-3916}, and
Fig.\,\ref{rxj1534.1-3916}. The orbital period is certainly much longer
than the 2574 days over which we have observed the star. Thus we can not
derive an orbit yet.  The eccentricity has to be very high ($\approx
0.9$).  \object{RXJ1534.1-3916} nicely illustrates the difficulty in
finding long-period spectroscopic binaries with eccentric orbits. If we
would not have taken the first two spectra, we probably would have
classified this object as a single star. According to Hogeveen
(\cite{hogeveen92}) only 0.45\% of the binaries have $e \geq 0.9$.

\begin{table}
\caption{RXJ 1534.1-3916}
\begin{tabular}{cr}
\hline \hline
HJD & RV [$km\,s^{-1}$]\\
\hline
245\,1261.76390 & $-25.0\pm0.1$ \\
245\,1335.77111 & $-27.5\pm1.0$ \\
245\,2093.73652 & $-12.7\pm1.5$ \\
245\,2372.74166 & $-10.6\pm0.1$ \\
245\,2384.71011 & $-10.9\pm0.5$ \\
245\,2395.72062 & $-11.4\pm0.4$ \\
245\,2710.78396 & $-12.3\pm0.5$ \\
245\,2717.78285 & $-12.0\pm0.2$ \\
245\,2723.77086 & $-11.3\pm0.2$ \\
245\,3096.73954 & $-12.4\pm0.7$ \\
245\,3104.77290 & $-11.3\pm0.4$ \\
245\,3105.73478 & $-11.5\pm0.5$ \\
245\,3122.76000 & $-13.0\pm0.1$ \\
245\,3137.77322 & $-12.6\pm0.4$ \\
245\,3168.62275 & $-11.6\pm0.1$ \\
245\,3501.73605 & $-13.7\pm0.1$ \\
245\,3523.74779 & $-13.8\pm0.1$ \\
245\,3556.77084 & $-11.3\pm0.2$ \\
245\,3556.78210 & $-11.1\pm0.5$ \\
245\,3601.55318 & $-13.6\pm0.3$ \\
245\,3603.51374 & $-10.8\pm0.2$ \\
245\,3811.68239 & $-12.0\pm0.2$ \\
245\,3813.84489 & $-12.1\pm0.7$ \\
245\,3835.23052 & $-11.9\pm0.3$ \\ 
\hline\hline
\end{tabular}
\label{RV_rxj1534.1-3916}
\end{table}

\subsection{RXJ1559.2-3814}

\begin{figure}[h]
\includegraphics[width=0.52\textwidth, angle=0]{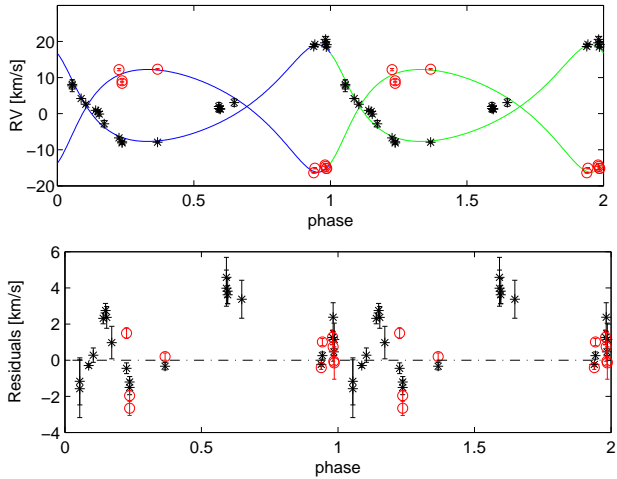}
\caption{The figure shows the RV-measurements of \object{RXJ 1559.2-3814}.
Shown is the orbit of the A and the B component for a period of 474
days.  }
\label{rxj1559.2-3814}
\end{figure}

\object{RXJ1559.2-3814} is a WTTS and an SB2. The average equivalent
width of H$\alpha$ is $-1.4\pm0.2$ \AA.  The equivalent width of the LiI
6708 \AA \, line is 0.23 \AA \, for the primary, and 0.14 \AA \, for the
secondary component.  We find that the ratio of the height of the peak of
the cross-correlation function of the two components is $1.9\pm0.2$.
Tab.\,\ref{RV_rxj1559.2-3814} gives the RV measurements obtained for
both components and the orbital elements are listed in the
Tab.\,\ref{orbit_rxj1559.2-3814}. Good fit is obtained for a period of
474 days (Fig.\,\ref{rxj1559.2-3814}).

\begin{table}
\caption{RXJ 1559.2-3814}
\begin{tabular}{crr}
\hline \hline
HJD & RV [$km\,s^{-1}$] & RV [$km\,s^{-1}$] \\
    & A component       & B component \\
\hline
245\,1264.71496 & $-6.7\pm0.3$ & $12.2\pm0.6$ \\ 
245\,1331.71643 & $-7.9\pm0.2$ & $12.3\pm1.1$ \\  
245\,1622.78442 & $19.6\pm0.4$ & $-14.2\pm0.4$ \\  
245\,1623.78926 & $20.6\pm0.8$ & $-14.6\pm0.6$ \\  
245\,1625.71327 & $19.1\pm0.9$ & $-15.2\pm0.7$  \\  
245\,3097.64088 & $4.2\pm0.1^1$  & \\
245\,3105.69324 & $2.6\pm0.4^1$  & \\
245\,3122.77331 & $0.9\pm0.3^1$  & \\
245\,3137.78841 & $-2.8\pm0.9$ & \\
245\,3168.81587 & $-8.1\pm0.4$ & $8.4\pm0.8$ \\ 
245\,3168.82963 & $-7.8\pm0.3$ & $9.1\pm1.0$ \\ 
245\,3501.75306 & $18.6\pm0.1$ & $-16.4\pm1.1$ \\
245\,3503.78331 & $19.2\pm0.2$ & $-15.1\pm0.4$ \\
245\,3523.76173 & $18.5\pm0.2$ & $-15.2\pm1.1$ \\
245\,3556.79982 & $ 7.7\pm1.6^1$ & \\
245\,3556.81105 & $ 8.1\pm1.3^1$ & \\
245\,3601.56957 & $ 0.6\pm0.4^1$ & \\
245\,3603.53176 & $-0.1\pm0.6^1$ & \\
245\,3811.70502 & $2.0\pm1.1^1$ & \\
245\,3811.72015 & $1.4\pm1.0^1$ & \\
245\,3813.81043 & $1.4\pm0.7^1$ & \\
245\,3813.82554 & $1.2\pm0.5^1$ & \\
245\,3838.28599 & $3.1\pm1.0^1$ & \\ 
\hline\hline
\end{tabular}
\label{RV_rxj1559.2-3814}
\hskip 2.0truecm $^1$ Just A-component.
\end{table}

\begin{table}
\caption{Orbital elements RXJ 1559.2-3814}
\begin{tabular}{ll}
\hline \hline
element & value \\
\hline
P           & $474.0\pm1.0$ $d$ \\
$T_0$ [HJD] & $2447834\pm3$ \\
$\gamma$    & $2.0 \pm0.4$ $km\,s^{-1}$ \\
$K_{1}$     & $13.4\pm0.2$ $km\,s^{-1}$ \\
$K_{2}$     & $14.2\pm0.2$ $km\,s^{-1}$ \\
e           & $ 0.336\pm0.005$ \\
$\omega_1$  & $336.8\pm1.9^o$ \\
$\omega_2$  & $156.8\pm1.9^o$ \\
$a_1 sin\,i$ & $0.549\pm 0.010 $ AU\\
$a_2 sin\,i$ & $0.581\pm 0.010$  AU\\
$q={{m_2}\over{m_1}}$ & $0.945 \pm 0.027$ \\ 
$m_1 sin^3\,i$ & $0.444 \pm 0.026$ M$_\odot$ \\
$m_2 sin^3\,i$  & $0.419 \pm 0.025$ M$_\odot$ \\
\hline\hline
\end{tabular}
\label{orbit_rxj1559.2-3814}
\end{table}


\subsection{GSC 06209-00735}

\begin{figure}[h]
\includegraphics[width=0.48\textwidth, angle=0]{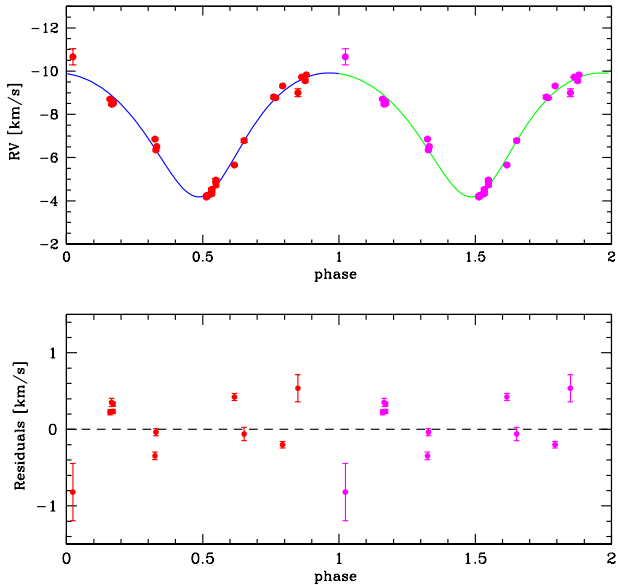}
\caption{The figure shows the phase-folded RV-measurements of
\object{GSC 06209-00735}. The orbital period is $2045\pm16$ days.}
\label{GSC6209-735}
\end{figure}

\object{GSC 06209-00735} was first identified as a member Upper Sco
association based on the spectroscopic and X-ray properties by Preibisch
et al. \cite{preibisch98}. We confirm the large equivalent width of the
LiI 6708 \AA \, line of $0.37\pm0.01\,AA $, and found that the object is a
SB1. H$\alpha$ is in absorption, the equivalent width is $0.3\pm0.1$
\AA .  In total, we took 26 spectra within 2287 days. We find an orbital
period of $2045 \pm16$ days, which is only slightly lower than the time
over which we observed the star. Fig.\,\ref{GSC6209-735} shows the
phase-folded RV-measurements together with the orbit.  The
RV-measurements are given in Tab.\,\ref{RV_GSC6209-735}, and the orbital
elements in Tab.\,\ref{orbit_GSC6209-735}.

\begin{table}
\caption{GSC 06209-00735}
\begin{tabular}{cr}
\hline \hline
HJD & RV [$km\,s^{-1}$] \\
\hline
245\,1260.79532 & $ -5.7\pm0.1$ \\
245\,1333.79536 & $ -6.7\pm0.1$ \\
245\,1622.87285 & $ -9.3\pm0.1$ \\
245\,1737.59186 & $-10.1\pm0.2$ \\
245\,2093.77154 & $-10.4\pm0.4$ \\
245\,2395.75208 & $ -8.6\pm0.1$ \\
245\,2396.70830 & $ -8.5\pm0.1$ \\
245\,2372.81128 & $ -8.7\pm0.1$ \\
245\,2384.73607 & $ -8.4\pm0.1$ \\
245\,2710.79456 & $ -6.8\pm0.1$ \\
245\,2717.79687 & $ -6.4\pm0.1$ \\
245\,2723.78469 & $ -6.5\pm0.1$ \\
245\,3096.76265 & $ -4.1\pm0.1$ \\
245\,3097.75000 & $ -4.2\pm0.1$ \\
245\,3104.79092 & $ -4.3\pm0.1$ \\
245\,3105.87514 & $ -4.3\pm0.1$ \\
245\,3122.78632 & $ -4.4\pm0.1$ \\
245\,3137.81404 & $ -4.5\pm0.1$ \\
245\,3168.63464 & $ -5.0\pm0.1$ \\
245\,3601.59793 & $ -8.8\pm0.1$ \\
245\,3616.52275 & $ -8.8\pm0.1$ \\
245\,3811.75370 & $ -9.7\pm0.1$ \\
245\,3838.36166 & $ -9.5\pm0.1$ \\ 
245\,3847.35257 & $ -9.8\pm0.1 $ \\
\hline\hline
\end{tabular}
\label{RV_GSC6209-735}
\end{table}

\begin{table}
\caption{Orbital elements of GSC 06209-00735}
\begin{tabular}{ll}
\hline \hline
element & value \\
\hline
P           & $2045\pm16$ $d$ \\
$T_0$ [HJD] & $2451022\pm12$ \\
$\gamma$       & $-7.62\pm0.02$ $km\,s^{-1}$ \\
$K_{1}$     & $2.87\pm0.05$ $km\,s^{-1}$ \\
e           & $0.20\pm0.03$ \\
$\omega$    & $8.1\pm2.3^o$ \\
$a_1 sin\,i$ & $0.53\pm0.018$ AU \\
f(m) & $ 0.0049\pm0.0005$ $M_\odot$\\
\hline\hline
\end{tabular}
\label{orbit_GSC6209-735}
\end{table}


\subsection{GSC 06213-00306}

\begin{figure}[h]
\includegraphics[width=0.48\textwidth, angle=0]{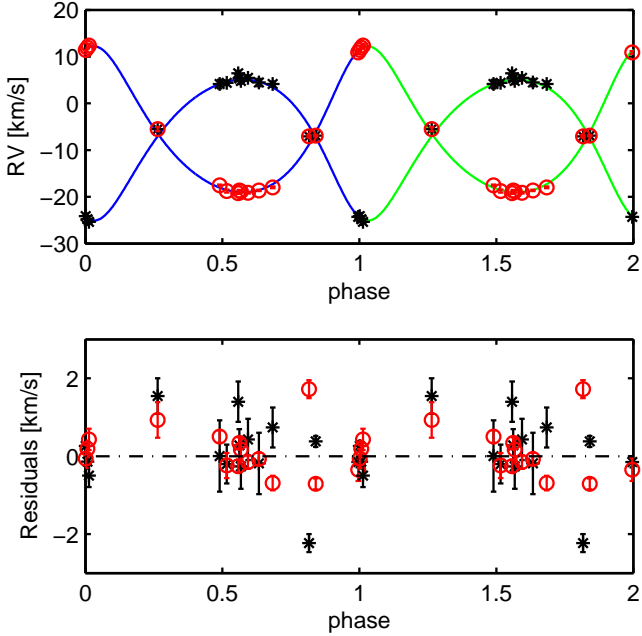}
\caption{The figure shows the phase-folded RV-measurements of 
\object{GSC 06213-00306}, together with the residuals. The orbital
period is $166.87\pm0.13$ days.}
\label{GSC06213-00306}
\end{figure}

\object{GSC 06213-00306} is another SB2 WTTS. The equivalent widths of
the Li\,I\,6708 \AA\,line are $0.241\pm0.004$ \AA , and $0.18\pm0.04$
\AA, for the primary and secondary, respectively. The H$\alpha$-line is
often filled in, occasionally it appears as an emission line with an
equivalent width of -0.1 \AA, at other occasions as an absorption line
with an equivalent width of 0.1 \AA. In total, we took 22 spectra of
this binary. We find an orbital solution with a period of $166.9\pm0.1$
days.  Thus, \object{GSC 06213-00306} is a long-period young SB2-binary,
the orbital elements for both components have been solved.  Given that
it has a K-magnitude of $7.43\pm0.02$ it is an ideal VLTI target.  The
RV-measurements are listed in Tab.\,\ref{RV_GSC06213-00306}, the orbit
in Tab.\,\ref{orbit_GSC06213-00306}.  The RV-measurements and the
orbit are shown in Fig.\,\ref {GSC06213-00306}.  It is interesting to
note that the masses of both components are almost identical.

\begin{table}
\caption{GSC 06213-00306}
\begin{tabular}{crr}
\hline \hline
HJD & RV [$km\,s^{-1}$] & RV [$km\,s^{-1}$] \\
    & A component       & B component       \\
\hline
245\,1333.82932 & $ -5.5\pm0.5^1$ & \\
245\,1622.89550 & $-24.3\pm0.1 $  & $10.9\pm0.3$ \\
245\,1623.75952 & $-24.1\pm0.1 $  & $11.4\pm0.2$ \\
245\,1624.73672 & $-24.8\pm0.2 $  & $11.9\pm0.2$ \\
245\,1625.73770 & $-25.3\pm0.3 $  & $12.4\pm0.3$ \\
245\,1737.66113 & $-18.0\pm0.2 $  & $ 4.1\pm0.5 $ \\
245\,2089.46697 & $ -9.0\pm0.2^1$ & \\ 
245\,2093.50599 & $ -7.0\pm0.2^1$ & \\ 
245\,2097.59641 & $ -6.9\pm0.1^1$ & \\
245\,2396.72380 & $  4.5\pm0.8$   & $-18.6\pm0.2$ \\
245\,2372.76696 & $  4.1\pm0.9$   & $-17.5\pm0.2$ \\ 
245\,2384.74816 & $  5.3\pm0.4$   & $-18.6\pm0.1$ \\ 
245\,2385.71382 & $  4.8\pm0.6$   & $-18.8\pm0.3$ \\ 
245\,2710.82314 & $  4.4\pm0.5$   & $-18.8\pm0.3$ \\ 
245\,2717.82525 & $  6.4\pm0.5$   & $-19.2\pm0.2$ \\ 
245\,2723.81319 & $  5.5\pm0.5$   & $-19.1\pm0.2$ \\
245\,1260.89577 & $ -6.0\pm0.4^1$ & \\
245\,1260.89577 & $ -6.1\pm0.5^1$ & \\
245\,3603.57132 & $  5.7\pm0.6$   & $-19.7\pm0.9$ \\
245\,3811.76438 & $ -2.6\pm1.0^2$   & \\
245\,3813.87632 & $ -2.8\pm1.0^2$   & \\
245\,3838.38271 & $ -7.0\pm0.4^2$   & \\ 
\hline\hline
\end{tabular}
\label{RV_GSC06213-00306}
\hskip 3.0truecm $^1$ both components, unresolved. \\
\hskip 1.0truecm $^2$ just A-component. 
\end{table}

\begin{table}
\caption{Orbital elements GSC 06213-00306} 
\begin{tabular}{ll}
\hline \hline
element & value \\
\hline
P           & $166.9\pm0.1$ $d$ \\
$T_0$ [HJD] & $2452124.2\pm1.1$ \\
$\gamma$    & $-6.76\pm0.06$ $km\,s^{-1}$ \\
$K_{1}$     & $15.1\pm0.1$ $km\,s^{-1}$ \\
$K_{2}$     & $15.65\pm0.09$ $km\,s^{-1}$ \\
e           & $0.226\pm0.006$ \\
$\omega_1$  & $161.8\pm2.2^o$ \\
$\omega_2$  & $341.8\pm2.2^o$ \\
$a_1 sin\,i$ & $0.226\pm0.002$ AU \\
$a_2 sin\,i$ & $0.234\pm0.002$ AU \\
$q={{m_2}\over{m_1}}$ & $0.97\pm0.01$ \\
$ m_1 \sin^3{i}$ & $0.246 \pm 0.006 $ $M_\odot$ \\
$ m_2 \sin^3{i}$ & $0.239 \pm 0.006 $  $M_\odot$\\
\hline\hline
\end{tabular}
\label{orbit_GSC06213-00306}
\end{table}



\section{Results: Spectroscopic binary candidates}

In this section we discuss two objects that were suspected to be
spectroscopic binaries of long period. We argue that in both
cases it is unlikely that these are spectroscopic binaries.

\subsection{VW Cha}

\object{VW Cha} was already observed by Melo (\cite{melo03}) with FEROS,
who found an occasional doubling of the cross-correlation peak and thus
suggested that this object is a spectroscopic binary. Brandner et
al. (\cite{brandner96}) found that \object{VW Cha} is a visual binary
with a separation of $0 \farcs 72$ $\pm$ $0 \farcs 03$ .  The companion
is only 0.25 mag fainter in J than the primary. Using K-band speckle
observations Ghez et al. (\cite{ghez97}) confirmed the presence of this
companion, and determine a separation of $0 \farcs 66 \pm 0 \farcs 03$.
Surprisingly, Ghez et al. \cite{ghez97} give a flux-ratio in the K-band
of $4.5\pm0.7$. Additionally to this inner component, Ghez et
al. (\cite{ghez97}) detected another component with a separation of 
$17\pm2''$.


As pointed out above, FEROS was operated up to October of 2002 (HJD
245\,2550) at the ESO 1.5-m telescope, and was then moved to the
MPG/ESO-2.20-m telescope. The entrance apertures were $2\farcs 7$ and $2
\farcs 0$ on the ESO 1.5-m telescope, and MPG/ESO 2.2-m telescope,
respectively.  Thus, if the fibre is centred on the primary, the
secondary would almost be at the edge.  The amount of light from the
primary and secondary thus would vary depending on the exact placement
of the fibre and the seeing.  The question thus arises whether the
occasional doubling of the spectral-lines is due to this effect, or
whether there is really another component.  If \object{VW Cha} were a
spectroscopic binary, its separation would have to be much smaller than
$0 \farcs 72$ $\pm$ $0 \farcs 03$, which implies that we should observe
large RV-variations.

In order to clarify the situation, we took 16 spectra of this star.  The
cross-correlation peak is in fact asymmetric. In contrast to our
expectation, we find only rather modest variations of the RV
(Tab.\,\ref{RV_VW_Cha}).  Additionally, the RV-measurements are erratic,
and not periodic.  While we can not completely rule out that this is an
SB2, it seems more likely that the asymmetric form of the
cross-correlation is due to the visual companion moving in and out of
the fibre, and not due to a spectroscopic companion. This
hypothesis would also explain the observed scatter of the
RV-measurements, because the velocity difference of two stars of one
solar-mass are in a circular orbit with a separtion of $0 \farcs 7$ is
about 4 $km\,s^{-1}$.  This object is in any case unsuitable for
VLTI-observations.


\begin{table}
\caption{VW Cha}
\begin{tabular}{cr}
\hline \hline
HJD & RV [$km\,s^{-1}$] \\
\hline
245\,1262.57801 & $14.9\pm0.5$ \\
245\,1333.57002 & $15.4\pm0.2$ \\
245\,2031.53243 & $17.2\pm0.2$ \\
245\,2319.70304 & $19.8\pm1.0$ \\
245\,2322.67626 & $18.6\pm0.6$ \\
245\,2372.53625 & $18.4\pm1.2$ \\
245\,2384.53498 & $15.3\pm0.4$ \\
245\,2710.59595 & $16.3\pm0.2$ \\
245\,2717.59503 & $16.3\pm0.4$ \\
245\,2723.58209 & $22.0\pm2.4$ \\
245\,3096.65990 & $18.7\pm0.9$ \\
245\,3104.58111 & $15.2\pm0.2$ \\
245\,3105.55467 & $17.5\pm0.1$ \\
245\,3122.59151 & $18.4\pm0.5$ \\
245\,3137.59880 & $16.1\pm0.4$ \\
245\,3155.64238 & $14.6\pm2.6$ \\
245\,3122.59151 & $18.5\pm0.4$ \\
245\,3522.57695 & $16.4\pm0.1$ \\
245\,3524.54647 & $18.3\pm0.4$ \\
\hline\hline
\end{tabular}
\label{RV_VW_Cha}
\end{table}

\subsection{GSC 06793-00569}
    
\object{GSC 06793-00569} is a WTTS with a weak H$\alpha$-emission line
of -0.4 \AA.  We have taken 19 spectra of this star. Although the
cross-correlation occasionally shows a blue asymmetry, and on other
occasions a red asymmetry, the RV-variations are moderate, and we do not
find any significant periodicity (Tab.\,\ref{RV_GSC6793-569}).  We thus
interpret the data in the same way as for \object{VW Cha}, the star could be
a visual binary with such a separation that roughly corresponds to the
radius of the fibre.

\begin{table}
\caption{GSC 06793-00569}
\begin{tabular}{cr}
\hline \hline
HJD & RV [$km\,s^{-1}$] \\
\hline
245\,1260.90498 & $-6.5\pm0.2$ \\
245\,1333.84131 & $-4.2\pm0.3$ \\
245\,1737.67610 & $-6.5\pm0.2$ \\
245\,2089.47704 & $-6.8\pm0.4$ \\
245\,2093.64727 & $-4.5\pm0.4$ \\
245\,2097.60976 & $-8.6\pm0.9$ \\
245\,2373.71233 & $-5.7\pm0.2$ \\
245\,2385.67462 & $-5.1\pm0.1$ \\
245\,2395.83337 & $-5.8\pm0.1$ \\
245\,2396.81929 & $-5.7\pm0.1$ \\
245\,2710.86108 & $-7.8\pm0.1$ \\
245\,2717.86133 & $-6.3\pm0.1$ \\
245\,2723.82421 & $-7.0\pm0.1$ \\
245\,3097.68383 & $-5.8\pm0.1$ \\
245\,3104.81253 & $-5.2\pm0.1$ \\
245\,3105.78900 & $-6.2\pm0.1$ \\
245\,3122.83664 & $-5.5\pm0.1$ \\
245\,3137.83685 & $-6.4\pm0.1$ \\
245\,3168.70120 & $-6.9\pm0.1$ \\
\hline\hline
\end{tabular}
\label{RV_GSC6793-569}
\end{table}

\section{Results: Single stars}

Unfortunately, not all previous surveys for pms-binaries give a
list of the objects that were found to be single. Such a list is however
very useful, because it avoids duplication of work.  In
Table\,\ref{tab:singleI}, \ref{tab:singleII}, and \ref{tab:singleIII} we
list the average RV-values obtained for all single stars together with
the variance of the RV-values determined for each star.  Also
given in these tables are the average equivalent widths of $H\alpha$ and
the Li\,I\,6708 \AA\,line, as derived from the spectra.  The K-band
brightness is taken from the 2MASS All-Sky Catalogue of Point Sources
(Cutri et al \cite{cutri03}). The number of spectra taken for each star
is given in the last column.

\begin{figure}[h]
\includegraphics[width=0.3\textwidth, angle=-90]{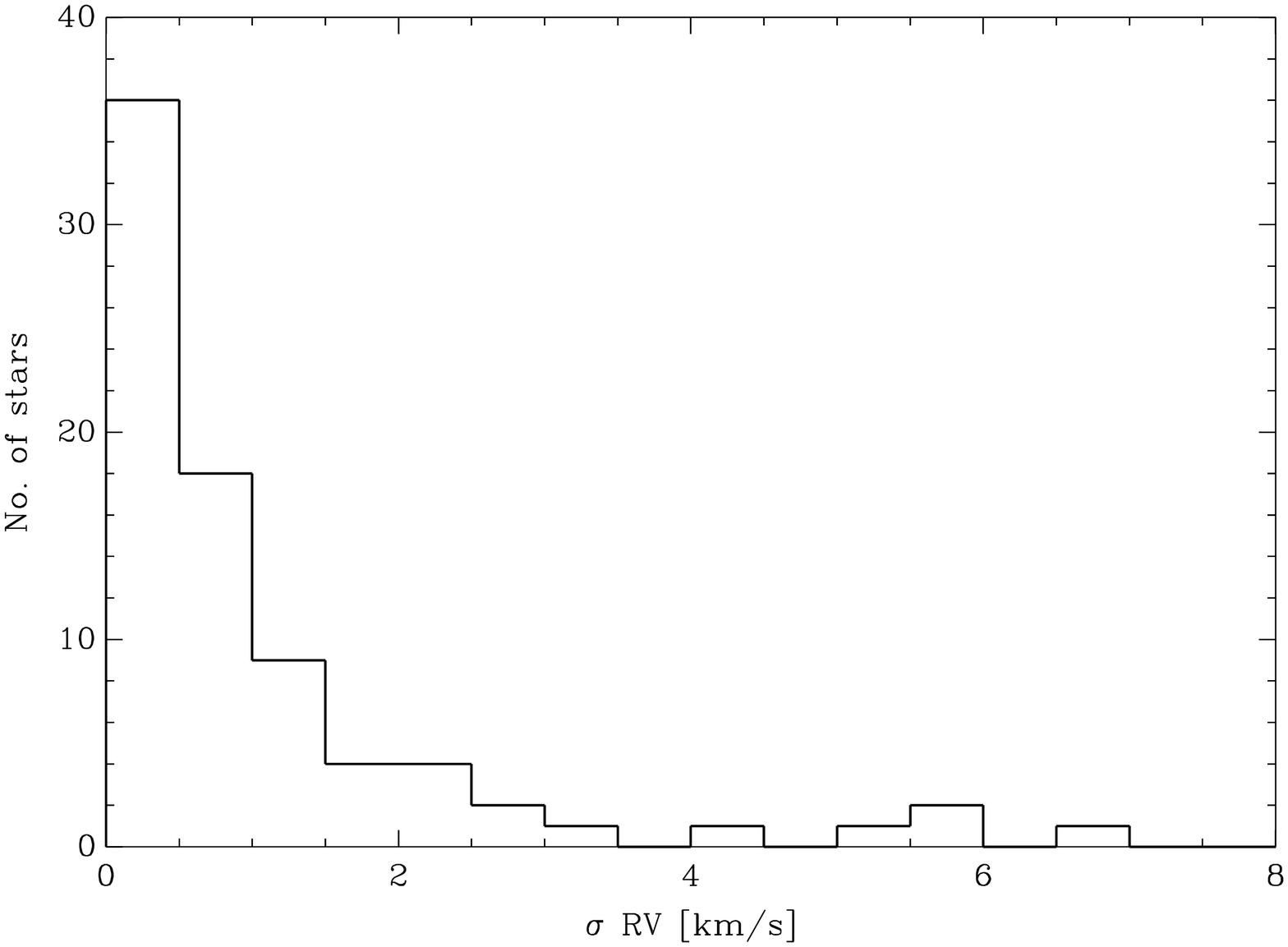}
\caption{Histogram of the RV-variations observed for 
the stars listed in Table\,\ref{tab:singleI}, \ref{tab:singleII}
\ref{tab:singleIII}. The amplitude of RV-variations caused are
typically less than 3 $km\,s^{-1}$.}
\label{SigmaRVHist}
\end{figure}

\begin{figure}[h]
\includegraphics[width=0.3\textwidth, angle=-90]{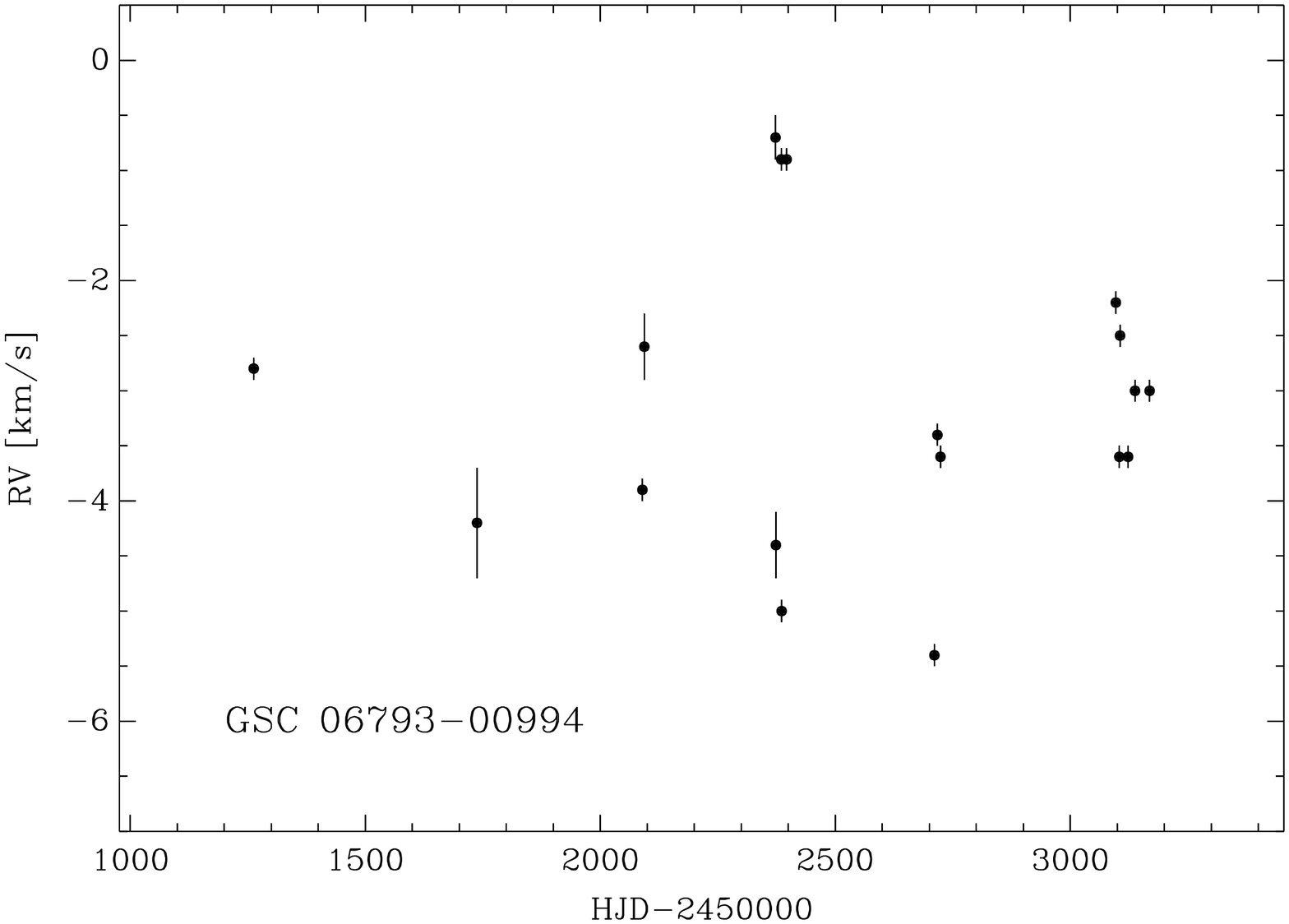}
\caption{A typical example for the non-periodic variations of a
young star. Shown are the RV-values of GSC 06793-00994.}
\label{RV_GSC06793-00994}
\end{figure}

The RV-curve of a star with a spot is periodic, where the period
corresponds to the rotation period of the star.  However, once this spot
vanishes and other spots appear at differnt longitudes, the amplitude
and the phase of the RV-signal changes. As a result, the periodic signal
vanishes, and the power in a periodogram decreases.  In contrast to
this, the signal of an orbiting companion remains unchaged and thus, the
power in a periodogram always increases, if more data is added.  The
differnce of the RV-signal caused by activity, and cuased by an orbiting
companion is nicely illustrated for \object{EK Dra}, $\epsilon$ Eri, and
\object{$\beta$ Gem}.  \object{EK Dra} is an active binary star with an
orbital period of 45 years, $\epsilon$ Eri is an active star with a
planet, and \object{$\beta$ Gem} an osciallting star with a planet
(K\"onig et al. \cite{koenig05}; Hatzes et al. \cite{hatzes00}; Hatzes
et al. \cite{hatzes06}). For more information on the effects of stellar
activity see also Paulson et al. \cite{paulson02}), and Saar et
al. \cite{saar98}.  Variations of the RV due to stellar activity have
already been observed in T Tauri stars. Because of the large activity
level of these stars, the effects are even larger than for older stars.
\object{RU Lup} for example, shows RV-variations with an amplitude of
2.17 $km\,s^{-1}$ (Stempels et al \cite {stempels07}). It is thus not
surpising that we observe the same effects in the T Tauri stars that we
observed.  Nice examples are \object{RXJ1415.0-7822}
(Tab.\,\ref{RV_RXJ1415.0-7822}), \object{HK Lup},
Tab.\,\ref{RV_HK_Lup}), \object{GSC 06793-01406}
Tab.\,\ref{RV_GSC6793-1406}), and \object{GSC 06793-00994} (Fig.\,\ref
{RV_GSC06793-00994}), of which we took 13 to 19 spectra each.  All of
these stars show RV-variations with a semi-amplitude of 1 to 3
$km\,s^{-1}$ but in none of them a significant period could be detected.
Fig.\,\ref{SigmaRVHist} shows a histogram of the RV-variations observed
in single stars.

The survey is limited to companions that cause a RV-amplitude of
$\geq$ 3 $km\,s^{-1}$.  For example, a system consisting of a 1.0
$M_\odot$ primary, and a 0.1 $M_\odot$ secondary would only be
detectable, if the period were equal, or shorter than a year. Similarly,
if the companion would have an orbital period of 3000 days, the mass of
the companion would have to be $\geq$ 0.23 $M_\odot$ in order to be
detectable.  The survey thus is in\-complete for long-period companions
with masses $\leq 0.2$ $M_\odot$.

We took 13 to 21 spectra of all those stars which show unusually
large RV-variations, in order to find out whether these are caused by
stellar activity, or by an orbiting companion. Apart from the objects
listed in Table\,\ref{tab:binariesI}, none of them turned out to be a
binary.  Only in the case of the two rapidly rotating stars \object{GSC
06781-01046} and \object{MN Lup}, we can not fully exclude that they are
binaries. One of the spectra of \object{GSC 06781-01046} shows a
double-line appearance. This object thus might be an SB2 but more
spectra are needed.

\begin{table}
\caption{RXJ1415.0-7822}
\begin{tabular}{cr}
\hline \hline
HJD & RV [$km\,s^{-1}$] \\
\hline
245\,1261.69317 & $19.7\pm0.4$ \\
245\,1622.67334 & $18.1\pm1.3$ \\
245\,1737.50348 & $17.1\pm1.7$ \\
245\,2019.67697 & $20.1\pm0.3$ \\
245\,2026.69069 & $19.0\pm0.6$ \\
245\,2031.67124 & $19.1\pm0.1$ \\
245\,2098.53621 & $16.2\pm1.0$ \\
245\,2372.67493 & $21.0\pm0.2$ \\
245\,2373.64078 & $19.3\pm0.2$ \\
245\,2384.64679 & $19.8\pm0.5$ \\
245\,2710.69367 & $19.2\pm0.3$ \\
245\,2717.71153 & $19.0\pm0.5$ \\
245\,2723.69779 & $18.9\pm0.4$ \\
245\,3096.86233 & $18.7\pm0.2$ \\
245\,3104.74397 & $20.3\pm0.5$ \\
245\,3105.70977 & $20.2\pm0.4$ \\
245\,3122.73806 & $18.6\pm0.5$ \\
245\,3137.75246 & $20.5\pm0.3$ \\
245\,3168.65138 & $18.7\pm0.3$ \\
\hline\hline
\end{tabular}
\label{RV_RXJ1415.0-7822}
\end{table}

\begin{table}
\caption{HK Lup}
\begin{tabular}{cr}
\hline \hline
HJD & RV [$km\,s^{-1}$] \\
\hline
245\,1261.90059 & $-1.3\pm0.5$ \\
245\,1262.69194 & $-5.0\pm0.5$ \\
245\,1622.81043 & $-5.3\pm0.7$ \\
245\,1733.68220 & $-2.5\pm0.2$ \\
245\,2026.83013 & $-5.2\pm0.7$ \\
245\,2031.79988 & $-5.7\pm1.2$ \\
245\,2093.75963 & $-1.4\pm0.4$ \\
245\,2098.59938 & $-1.9\pm0.3$ \\
245\,2395.79978 & $ 0.2\pm0.5$ \\
245\,2396.77110 & $ 4.5\pm0.2$ \\
245\,2372.75478 & $-0.8\pm0.4$ \\
245\,2384.72338 & $-2.3\pm0.1$ \\
245\,2385.69702 & $-0.6\pm0.1$ \\
245\,2710.80837 & $-0.6\pm0.4$ \\
245\,2717.81062 & $-2.3\pm0.1$ \\
245\,2723.79847 & $-0.4\pm0.1$ \\
245\,3096.77689 & $-1.8\pm0.3$ \\
245\,3104.80132 & $-2.3\pm2.9$ \\
245\,3105.76881 & $-8.6\pm0.4$ \\
245\,3122.81946 & $-8.7\pm0.2$ \\
245\,3455.75212 & $ 1.6\pm0.8$ \\
\hline\hline
\end{tabular}
\label{RV_HK_Lup}
\end{table}

\begin{table}
\caption{GSC 06793-01406}
\begin{tabular}{cr}
\hline \hline
HJD & RV [$km\,s^{-1}$] \\
\hline
245\,1625.80168 & $-6.9\pm0.6$ \\
245\,1737.70415 & $-7.4\pm0.6$ \\
245\,2089.73677 & $-7.3\pm0.4$ \\
245\,2093.78125 & $-5.8\pm0.7$ \\
245\,2384.81131 & $-5.5\pm0.1$ \\
245\,2717.88212 & $-7.9\pm0.2$ \\
245\,2723.87107 & $-6.8\pm0.1$ \\
245\,3096.84966 & $-5.7\pm0.3$ \\
245\,3104.85577 & $-5.6\pm0.1$ \\
245\,3105.83868 & $-7.1\pm0.1$ \\
245\,3122.87716 & $-7.0\pm0.2$ \\
245\,3137.88922 & $-8.0\pm0.2$ \\
245\,3168.74239 & $-6.2\pm0.1$ \\
\hline\hline
\end{tabular}
\label{RV_GSC6793-1406}
\end{table}

\begin{table}
\caption{GSC 06793-00994}
\begin{tabular}{cr}
\hline \hline
HJD & RV [$km\,s^{-1}$] \\
\hline
245\,1262.70144 & $-2.8\pm0.1$ \\
245\,1737.69110 & $-4.2\pm0.5$ \\
245\,2089.51975 & $-3.9\pm0.1$ \\
245\,2093.68633 & $-2.6\pm0.3$ \\
245\,2396.80296 & $-0.9\pm0.1$ \\
245\,2372.82468 & $-0.7\pm0.2$ \\
245\,2373.69694 & $-4.4\pm0.3$ \\
245\,2384.79676 & $-0.9\pm0.1$ \\
245\,2385.73404 & $-5.0\pm0.1$ \\
245\,2710.87369 & $-5.4\pm0.1$ \\
245\,2717.83773 & $-3.4\pm0.1$ \\
245\,2723.83505 & $-3.6\pm0.1$ \\
245\,3096.79173 & $-2.2\pm0.1$ \\
245\,3104.83363 & $-3.6\pm0.1$ \\
245\,3105.80278 & $-2.5\pm0.1$ \\
245\,3122.84741 & $-3.6\pm0.1$ \\
245\,3137.84760 & $-3.0\pm0.1$ \\
245\,3168.72226 & $-3.0\pm0.1$ \\
\hline\hline
\end{tabular}
\label{RV_GSC6793-994}
\end{table}



\begin{table*}
\caption{The single stars: Chamaeleon}
\begin{tabular}{llccccccrrl}
\hline \hline
            & region & EW H$\alpha^1$ & EW LiI & spec & $m_K$ & RA       & Dec      & RV           & $\sigma$ RV  & No. of \\
            &        & [\AA]        & [\AA]  & type & [mag] & (2000.0) & (2000.0) & [$kms^{-1}$] & [$kms^{-1}$] & spectra \\
\hline
EG Cha          & Cha & $-1.1\pm0.1$ & $0.53\pm0.01$ & K7 & $7.338\pm0.021$ & 08 36 56.2 & -78 56 46 & $17.4\pm0.2$ & 0.4 & 3\\  
EO Cha          & Cha & $-1.0\pm0.2$ & $0.53\pm0.04$ & M0 & $8.732\pm0.021$ & 08 44 31.9 & -78 46 31 & $17.1\pm0.3$ & 0.5 & 2 \\    
EQ Cha$^3$      & Cha & $-6.5\pm1.1$ & $0.57\pm0.04$ & M3 & $8.410\pm0.031$ & 08 47 59.2 & -78 54 58 & $17.9\pm1.0$ & 2.7 & 8 \\    
RXJ0850.1-7554  & Cha &    abs       & $0.27\pm0.01$ & G5 & $8.704\pm0.019$ & 08 50 05.4 & -75 54 38 & $18.1\pm1.0$ & 1.3 & 2 \\    
RXJ0915.5-7609  & Cha & $-1.1\pm0.2$ & $0.55\pm0.03$ & K6 & $8.488\pm0.033$ & 09 15 29.1 & -76 08 47 & $20.3\pm0.1$ & 0.1 & 3 \\     
RXJ0917.2-7744  & Cha &    abs       & $0.33\pm0.06$ & G2 & $8.812\pm0.023$ & 09 17 11.4 & -77 44 15 &  $5.7\pm1.8$ & 3.2 & 3 \\    
HD84075         & Cha &    abs       & $0.17\pm0.01$ & G2 & $7.160\pm0.016$ & 09 36 17.8 & -78 20 42 &  $5.2\pm0.1$ & 0.1 & 3 \\ 
HD86356         & Cha &     fi       & $0.34\pm0.05$ & G8 & $8.040\pm0.029$ & 09 51 50.7 & -79 01 38 & $13.2\pm0.4$ & 0.5 & 2 \\ 
HD86588         & Cha &    abs       & $0.11\pm0.01$ & F6 & $7.994\pm0.029$ & 09 53 13.7 & -79 33 28 & $ 2.4\pm0.5$ & 2.4 & 3 \\
RXJ1001.1-7913  & Cha & $-2.5\pm0.7$ & $0.10\pm0.01$ & K8 & $9.216\pm0.021$ & 10 01 08.8 & -79 13 08 & $14.4\pm0.4$ & 0.6 & 2 \\    
RXJ1005.3-7749  & Cha & $-4.2\pm0.1$ & $0.58\pm0.01$ & M1 & $8.892\pm0.019$ & 10 05 17.6 & -77 49 06 & $16.3\pm0.2$ & 0.3 & 2 \\    
SX Cha E$^4$    & Cha & $-20\pm5$    & $0.65\pm0.04$ & M0 & $8.685\pm0.024$ & 10 55 59.8 & -77 24 40 & $13.9\pm0.5$ & 1.1 & 5 \\
SY Cha          & Cha & $-13\pm2$    & $0.56\pm0.03$ & M0 & $8.631\pm0.019$ & 10 56 30.5 & -77 11 39 & $14.0\pm0.3$ & 0.5 & 3 \\
TW Cha          & Cha & $-28\pm11$   & $0.41\pm0.01$ & K7 & $8.616\pm0.021$ & 10 59 01.1 & -77 22 41 & $17.8\pm0.1$ &  0.1 & 2 \\
CR Cha (=Sz6)   & Cha & $-34\pm2$    & $0.43\pm0.04$ & K2 & $7.310\pm0.023$ & 10 59 07.0 & -77 01 40 & $16.7\pm1.1$ & 2.0 & 3 \\
CT Cha          & Cha & $-40\pm21$   & $0.40\pm0.05$ & K7 & $8.661\pm0.021$ & 11 04 09.1 & -76 27 19 & $15.2\pm0.3$ & 0.9 & 13 \\
DI Cha$^5$      & Cha & $-17\pm1$    & $0.26\pm0.01$ & G2 & $6.217\pm0.020$ & 11 07 20.7 & -77 38 07 & $14.6\pm0.2$ & 0.3 & 2 \\
VW Cha$^6$      & Cha & $-57\pm11$   & $0.44\pm0.05$ & M0.5 & $6.962\pm0.026$ & 11 08 01.8 & -77 42 29 & $17.2\pm0.5$ & 2.0 & 17 \\ 
WW Cha$^2$      & Cha & $-57\pm13$   & $0.3\pm0.2$ & K5 & $6.083\pm0.053$ & 11 10 00.7 & -76 34 59 & $15.0\pm2.0$ & 0.5 & 2 \\
HBC 584         & Cha & $-77\pm1$    & $0.3\pm0.1$ & K7 & $9.175\pm0.024$ & 11 10 49.6 & -77 17 52 & $13.6\pm0.1$ & 0.2 & 2 \\
Sz 41$^7$       & Cha & $-2.0\pm0.9$ & $0.43\pm0.02$ & K0 & $7.999\pm0.031$ & 11 12 24.5 & -76 37 06 & $14.6\pm0.3$ &  1.1 & 18 \\
VZ Cha$^2$      & Cha & $-32\pm4$    & $0.31\pm0.01$ & K6 & $8.242\pm0.038$ & 11 09 23.8 & -76 23 21 & $16.3\pm0.4$ & 0.6 & 2 \\
CV Cha$^2,5$    & Cha & $-67\pm16$   & $0.34\pm0.02$ & G8 & $6.845\pm0.026$ & 11 12 27.7 & -76 44 22 & $16.1\pm1.2$ & 1.8 & 2 \\
GSC 07739-02180 & SC  & $-3.9\pm1.2$ & $0.52\pm0.01$ &    & $8.053\pm0.029$ & 11 21 05.5 & -38 45 17 & $12.1\pm0.7$ & 1.2 & 3 \\   
RXJ1129.2-7546  & Cha &    fi        & $0.49\pm0.02$ & K3 & $8.878\pm0.024$ & 11 29 12.7 & -75 46 26 & $14.6\pm0.5$ & 0.8 & 3 \\    
RXJ1140.3-8321  & Cha & $-0.3\pm0.2$ & $0.21\pm0.01$ & K2 & $8.635\pm0.019$ & 11 40 16.6 & -83 21 00 & $12.7\pm0.1$ & 0.1 & 3 \\    
RXJ1150.4-7704  & Cha & $-1.3\pm0.5$ & $0.40\pm0.06$ & K2 & $8.970\pm0.021$ & 11 50 28.9 & -77 04 38 & $-3.3\pm1.0$ & 2.1 & 4 \\    
T Cha           & Cha & $-8.8\pm9.9$ & $0.41\pm0.03$ & G8 & $6.954\pm0.018$ & 11 57 13.5 & -79 21 32 & $14.0\pm1.3$ & 5.3 & 17 \\
RXJ1159.7-7601  & Cha & $-0.2\pm0.1$ & $0.46\pm0.01$ & K2 & $8.304\pm0.027$ & 11 59 42.3 & -76 01 26 & $13.8\pm0.1$ & 0.1 & 3 \\  
RXJ1202.1-7853  & Cha & $-3.3\pm0.5$ & $0.56\pm0.02$ & K7 & $8.307\pm0.023$ & 12 02 03.8 & -78 53 01 & $17.1\pm0.2$ & 0.3 & 3 \\  
HD106772        & Cha &  abs         & $0.21\pm0.01$ & G2 & $6.231\pm0.031$ & 12 17 26.9 & -80 35 06 & $-12.9\pm0.2$ & 0.3 & 2 \\  
RXJ1219.7-7403  & Cha & $-4.5\pm1.6$ & $0.56\pm0.01$ & K8 & $8.858\pm0.023$ & 12 19 43.8 & -74 03 57 & $13.8\pm0.1$ & 0.2 & 2 \\  
RXJ1220.4-7407  & Cha & $-2.9\pm0.6$ & $0.61\pm0.01$ & K7 & $8.367\pm0.022$ & 12 20 21.9 & -74 07 39 & $12.3\pm0.4$ & 0.6 & 2 \\  
HD107722        & Cha &   abs        & $0.08\pm0.01$ & F6 & $7.135\pm0.034$ & 12 23 29.0 & -77 40 51 & $13.4\pm0.4$ & 0.6 & 2 \\
RXJ1233.5-7523  & Cha &   abs        & $0.14\pm0.01$ & K0 & $7.756\pm0.040$ & 12 33 32.0 & -75 23 25 & $15.5\pm0.1$ & 0.1 & 2 \\  
RXJ1239.4-7502  & Cha &   abs        & $0.41\pm0.01$ & K2 & $7.777\pm0.021$ & 12 39 21.3 & -75 02 39 & $13.9\pm0.2$ & 0.3 & 2 \\  
BC Cha          & Cha & $-72\pm18$   & $0.52\pm0.03$ &    & $9.354\pm0.021$ & 13 01 59.0 & -77 51 22 & $14.3\pm0.2$ & 0.4 & 2 \\
RXJ1325.7-7955  & Cha &   abs        & $0.10\pm0.05$ & K1 & $8.932\pm0.022$ & 13 25 42.8 & -79 55 25 & $-8.4\pm3.6$ & 7.1 & 4 \\  
CPD-75~902      & Cha &   abs        & $0.22\pm0.01$ & K0 & $8.015\pm0.053$ & 13 49 12.9 & -75 49 48 & $-0.8\pm0.1$ & 0.2 & 2 \\
RXJ1415.0-7822  & Cha &   fi         & $0.26\pm0.05$ & G5 & $9.876\pm0.023$ & 14 15 01.8 & -78 22 12 & $-3.1\pm0.3$ & 1.3 & 18 \\
\hline\hline
\end{tabular}
\label{tab:singleI}
$^1$ em=emission, abs=absorption fi=filled in \\
$^2$ WW Cha, VZ Cha, CV Cha: also observed by Melo et al. (\cite{melo03}) and found to be single \\ 
$^3$ EQ Cha : difficult object, cc-function asymmetric, unlikely to be binary \\
$^4$ SX Cha : visual binary M0 primary and M3 secondary with a separation of $2 \farcs 2$. 
             (Reipurth \& Zinnecker \cite{reipurth93}), only eastern component observed \\
$^5$ DI Cha (LkH$\alpha$ 332-17): binary with separation of $4 \farcs 9\pm 0\farcs 2$ (Ghez al. \cite{ghez97}) \\
$^6$ VW Cha : triple system, separation of $0 \farcs 66\pm 0 \farcs 03$ and $17 \farcs \pm 2\farcs $ (Ghez et al. \cite{ghez97}) \\
$^7$ Sz 41: triple system, separation of $1 \farcs 5\pm0 \farcs 8$ and $12 \farcs 4\pm0 \farcs 6$ (Ghez et al. \cite{ghez97})  \\
\end{table*} 

\begin{table*}
\caption{The single stars: Lupus, Scorpius Centaurus}
\begin{tabular}{llccccccrrl}
\hline \hline
            & region & EW H$\alpha^1$ & EW LiI & spec & $m_K$ & RA       & Dec      & RV           & $\sigma$ RV  & No. of \\
            &        & [\AA]        & [\AA]  & type & [mag] & (2000.0) & (2000.0) & [$kms^{-1}$] & [$kms^{-1}$] & spectra \\
\hline
LQ Lup          & Lup & $-1.5\pm0.5$ & $0.26\pm0.03$ & G8 & $8.809\pm0.021$ & 15 08 37.8 & -44 23 17 & $7.6\pm1.5$ & 5.8 & 17 \\
LS Lup          & Lup & $-0.7\pm0.1$ & $0.47\pm0.01$ & K1 & $9.454\pm0.021$ & 15 15 52.8 & -44 18 16 & $6.8\pm0.4$ & 0.5 & 2 \\
RXJ1516.6-4406  & Lup &     fi       & $0.54\pm0.13$ & K2 & $9.193\pm0.019$ & 15 16 36.8 & -44 07 19 & $4.4\pm1.3$ & 5.5 & 18 \\
MM Lup          & Lup &     fi       & $0.47\pm0.01$ &    & $9.260\pm0.023$ & 15 23 25.7 & -40 55 45 & $4.4\pm0.1$ & 0.2 & 3 \\
MN Lup$^{3}$   & Lup & $-6.1\pm0.8$ & $0.44\pm0.05$ & M2 & $9.496\pm0.019$ & 15 23 30.4 & -38 21 29 & $6.6\pm3.4$ & 6.9 & 4 \\
MP Lup          & Lup & fi$^1$       & $0.36\pm0.01$ & K1 & $8.930\pm0.019$ & 15 24 32.4 & -36 52 02 & $3.3\pm0.3$ & 0.6 & 4 \\
MQ Lup          & Lup & $-0.5\pm0.2$ & $0.39\pm0.03$ & K2 & $8.842\pm0.021$ & 15 25 33.2 & -36 13 46 & $3.6\pm0.1 $ & 0.1 & 2 \\
MR Lup          & Lup & $-1.7\pm0.2$ & $0.47\pm0.03$ & K6 & $8.963\pm0.019$ & 15 25 36.7 & -35 37 32 & $-1.0\pm1.9$ & 4.2 & 5 \\
HIP75836$^{4}$ &    & abs          & $0.04\pm0.01$ & K0 & $6.760\pm0.020$ & 15 29 26.9 & -28 50 52 & $-19.5\pm0.1$ & 0.1 & 3 \\  
MT Lup          & Lup & $-1.3\pm0.5$ & $0.48\pm0.01$ & K5 & $9.376\pm0.024$ & 15 38 02.7 & -38 07 22 & $3.0\pm0.2 $ & 0.3 & 2 \\
RXJ1538.6-3916  & Lup & abs          & $0.39\pm0.01$ & K4 & $8.854\pm0.023$ & 15 38 38.3 & -39 16 54 & $2.5\pm0.2$ & 0.4 & 3 \\
RXJ1539.2-4455  & Lup & $-1.8\pm0.4$ & $0.17\pm0.03$ &    & $9.989\pm0.021$ & 15 39 12.0 & -44 55 29 & $5.5\pm0.1$ & 0.1 & 2 \\
MU Lup          & Lup & $-0.8\pm0.1$ & $0.49\pm0.01$ & K6 & $9.187\pm0.026$ & 15 40 41.2 & -37 56 18 & $2.2\pm0.1$ & 0.2 & 2 \\
GSC 06785-00476 & SC  &  abs         & $0.30\pm0.01$ & G7 & $8.920\pm0.023$ & 15 41 06.8 & -26 56 26 & $-2.0\pm0.1$ & 0.2 & 3 \\   
GSC 06781-01046$^{5}$ & SC  &  abs         & $\leq0.2$     & G5 & $8.177\pm0.020$ & 15 42 49.9 & -25 36 41 &  SB2? & - & 2 \\   
GW Lup$^2$      & Lup & $-52.8\pm6.8$ & $0.48\pm0.04$ & M0 & $8.630\pm0.021$ & 15 46 44.7 & -34 30 36 & $-3.5\pm0.5$ & 0.8 & 2 \\
HBC 603$^{6}$   & Lup & $-10.4\pm2.2$ & $0.58\pm0.01$ & M0 & $8.271\pm0.026$ & 15 51 47.0 & -35 56 43 & $-2.6\pm0.1$ & 0.2 & 3 \\
RXJ1555.4-3338  & Lup & $-0.7\pm0.5$ & $0.44\pm0.01$ & K5 & $9.353\pm0.022$ & 15 55 26.3 & -33 38 22 & $0.0\pm0.1 $ & 0.2 & 3 \\
HD142987        & SC  & $-2.1\pm0.1$ & $\sim 0.3$    & G3 & $7.614\pm0.021$ & 15 58 20.6 & -18 37 25 & $6.0\pm1.5$ & 2.2 & 2 \\   
GSC 06191-00552 & SC  & $-0.6\pm0.1$ & $0.48\pm0.01$ & K3 & $8.325\pm0.024$ & 15 58 47.8 & -17 57 59 & $-6.4\pm0.5$ & 0.9 & 3 \\      
GSC 06195-00768 & SC  & $-0.5\pm0.2$ & $0.46\pm0.02$ & K7 & $8.372\pm0.025$ & 15 57 02.4 & -19 50 41 & $-5.1\pm0.2$ & 0.5 & 4 \\   
HBC 609         & Lup & $-30\pm1.0$  & $0.39\pm0.06$ & K8 & $8.608\pm0.023$ & 15 59 16.5 & -41 57 09 & $0.2\pm0.7 $ & 1.0 & 2 \\
RY Lup          & Lup & $-3.2\pm0.3$ & $0.38\pm0.01$ & G0 & $6.976\pm0.019$ & 15 59 28.4 & -40 21 51 & $-0.4\pm0.5$ & 0.8 & 2 \\
MZ Lup          & Lup &   abs        & $0.33\pm0.02$ & G8 & $8.528\pm0.030$ & 16 01 09.0 & -33 20 14 & $1.8\pm0.2 $ & 0.3 & 2 \\
GSC 06204-00812 & SC  & $-0.1\pm0.1$ & $0.47\pm0.01$ & K4 & $8.727\pm0.025$ & 16 03 02.7 & -18 06 04 & $-4.9\pm0.1 $ & 0.1 & 2 \\   
GSC 06784-01219 & SC  & $-0.1\pm0.1$ & $0.37\pm0.01$ & G7 & $8.461\pm0.023$ & 16 05 50.5 & -25 33 12 & $-4.2\pm0.1$ & 0.2 & 2 \\   
EX Lup$^2$      & Lup & $-24\pm8$    & $0.36\pm0.08$ & M0 & $8.496\pm0.021$ & 16 03 05.5 & -40 18 25 & $-0.3\pm0.1$ & 0.3 & 3 \\
HBC 613         & Lup & $-32\pm2$    & $0.46\pm0.02$ & K8 & $8.724\pm0.023$ & 16 07 10.0 & -39 11 03 & $-0.5\pm0.3$ & 0.5 & 2 \\
NQ Lup          & Lup & $-1.4\pm0.1$ & $0.51\pm0.03$ & K7 & $8.714\pm0.021$ & 16 08 18.3 & -38 44 04 & $0.8\pm0.1$ & 0.1 & 2 \\
GSC 06784-00039 & SC  &  fi          & $0.37\pm0.01$ & G7 & $7.908\pm0.016$  & 16 08 43.4 & -26 02 17 & $-4.6\pm0.5$ & 0.7 & 2 \\   
HK Lup$^2$      & Lup & $-22\pm17$   & $0.54\pm0.03$ & K8 & $8.014\pm0.021$ & 16 08 22.5 & -39 04 46 & $-2.2\pm0.8$ & 2.8 & 21 \\
V1094 Sco       & Lup & $-8.2\pm3.3$ & $0.49\pm0.02$ & K6 & $8.658\pm0.021$ & 16 08 36.2 & -39 23 03 & $0.4\pm0.3$ & 1.0 & 9 \\
GSC 06793-00868 & SC  & $-2.8\pm0.2$ & $0.55\pm0.01$ & M1 & $8.815\pm0.034$ & 16 11 56.4 & -23 04 04 & $-6.0\pm0.8$ & 1.3 & 3 \\   
GSC 06793-00797 & SC  & $-0.7\pm0.2$ & $0.52\pm0.01$ & K4 & $8.455\pm0.027$ & 16 13 02.8 & -22 57 43 & $-6.5\pm0.4$ & 0.8 & 4 \\   
GSC 06793-00569$^7$& SC & $-0.4\pm0.2$ & $0.47\pm0.20$ & K1 & $8.494\pm0.019$ & 16 13 29.3 & -23 11 06 & $-6.1\pm0.2$ & 1.0 & 19 \\
GSC 06793-00994 & SC  & $-0.3\pm0.2$ & $0.38\pm0.01$ & G4 & $8.608\pm0.023$ & 16 14 02.1 & -23 01 01 & $-3.1\pm0.3$ & 1.3 & 18 \\
GSC 06801-00186 & SC  & abs          & $0.35\pm0.02$ & G5 & $8.686\pm0.022$ & 16 14 59.3 & -27 50 22 & $-1.7\pm0.1$ & 0.3 & 4 \\   
GSC 06793-01406 & SC  & $-0.6\pm0.2$ & $0.36\pm0.01$ & G7 & $8.102\pm0.018$ & 16 16 18.0 & -23 39 47 & $-6.7\pm0.2$ & 0.8 & 13 \\
GSC 06214-02384 & SC  & fi           & $0.41\pm0.01$ & K0 & $8.509\pm0.019$ & 16 19 34.0 & -22 28 29 & $-3.5\pm0.1$ & 0.1 & 3 \\   
GSC 06794-00156 & SC  & $-0.3\pm0.1$ & $0.34\pm0.01$ & G6 & $7.084\pm0.018$ & 16 24 51.4 & -22 39 32 & $-3.7\pm0.7$ & 1.3 & 4 \\   
GSC 06794-00537 & SC  & fi           & $0.48\pm0.02$ & K2 & $8.184\pm0.024$ & 16 23 07.8 & -23 00 59 & $-6.5\pm0.6$ & 1.0 & 3 \\   
GSC 06798-00035 & SC  & fi           & $0.35\pm0.02$ & G1 & $7.695\pm0.023$ & 16 23 32.3 & -25 23 48 & $-7.1\pm0.3$ & 0.2 & 2 \\
GSC 06794-00337 & SC  & fi           & $0.44\pm0.01$ & K1 & $8.084\pm0.026$ & 16 27 39.6 & -22 45 22 & $-6.1\pm0.1$ & 0.1 & 4 \\   
\hline\hline
\end{tabular}
\label{tab:singleII}
$^1$ em=emission, abs=absorption fi=filled in \\
$^2$ GW Lup, EX Lup, HK Lup : also observed by Melo et al. (\cite{melo03}) and found to be single \\ 
$^3$ MN Lup: large $v\,sin\,i$ possibly a spectroscopic binary but data insufficient \\
$^4$ HIP75836 : listed in Simbad as eclipsing binary but no RV-variations observed, unlikely to be member of Lup \\
$^5$ GSC 06781-01046 : this object might be a short-period SB2 \\
$^6$ HBC 603 (Sz77) : binary with separation of $1 \farcs 8\pm0 \farcs 1$  (Ghez et al. \cite{ghez97})
$^7$ GSC 06793-00569 : asymmetric cross-correlation function, could be a visual binary \\
\end{table*}

\begin{table*}
\caption{The single stars: $\rho$ Ophiuchi, Corona Australis}
\begin{tabular}{llccccccrrl}
\hline \hline
            & region & EW H$\alpha^1$ & EW LiI & spec & $m_K$ & RA       & Dec      & RV           & $\sigma$ RV  & No. of \\
            &        & [\AA]        & [\AA]  & type & [mag] & (2000.0) & (2000.0) & [$kms^{-1}$] & [$kms^{-1}$] & spectra \\
\hline
RXJ1612.3-1909     & Oph & $-3.9\pm0.1$ & $\leq 1.0$    & M2.5 & $9.605\pm0.019$ & 16 12 20.9   & -19 09 04 & $-1.6\pm1.6$ & 2.2 & 2 \\
V1002 Sco          & Oph &      fi      & $0.48\pm0.05$ & K0   & $7.494\pm0.026$ & 16 12 40.5 & -18 59 28 & $-3.7\pm1.0$ & 1.4 & 2 \\
V2503 Oph$^{2,3}$  & Oph & $-57\pm1$    & $0.50\pm0.03$ & K6-7 & $7.509\pm0.031$ & 16 25 10.5 & -23 19 14 & $-7.1\pm0.4$ & 0.6 & 2 \\
RXJ1625.3-2402     & Oph &      f1      & $0.52\pm0.12$ & K5   & $8.764\pm0.025$ & 16 25 22.4 & -24 02 06 & $-4.8\pm0.3$ & 0.4 & 2 \\
V2058 Oph          & Oph & $-91\pm12$   & $0.45\pm0.04$ & K5   & $7.518\pm0.024$ & 16 25 56.2 & -24 20 48 & $-4.9\pm0.3$ & 0.7 & 3 \\
V2129 Oph$^{2,4}$  & Oph & $-22\pm4$    & $0.53\pm0.02$ & K3.5 & $7.207\pm0.023$ & 16 27 40.3 & -24 22 03 & $-6.8\pm0.6$ & 1.7 & 8 \\
HBC 268  $^2$      & Oph & $-42\pm2$    & $0.44\pm0.02$ & K2-3 & $7.610\pm0.024$ & 16 31 33.5 & -24 27 37 & $-6.3\pm0.5$ & 0.9 & 4 \\
HD173148$^5$       & CrA & abs          & $0.27\pm0.03$ & G5   & $7.204\pm0.020$ & 18 45 34.8 & -37 50 20 & $0.1\pm0.4$ & 1.6 & 17 \\
S CrA $^{2,6}$     & CrA & $-51\pm6$    & $0.20\pm0.01$ & K6   & $6.107\pm0.021$ & 19 01 08.6 & -36 57 20 & $0.9\pm0.9$ & 1.6 & 3 \\
V709 CrA           & CrA & fi           & $0.40\pm0.02$ & K0   & $7.713\pm0.021$ & 19 01 34.9 & -37 00 57 & $-1.3\pm0.3$ & 0.7 & 6 \\
DG CrA             & CrA & $-53\pm16$   & $0.36\pm0.03$ &      & $7.952\pm0.026$ & 19 01 55.2 & -37 23 41 & $-1.8\pm0.4$ & 0.9 & 6 \\
V702 CrA           & CrA & abs,fi       & $0.32\pm0.01$ & G5   & $8.354\pm0.027$ & 19 02 02.0 & -37 07 44 & $-0.7\pm0.1$ & 0.5 & 19 \\
HBC 679            & CrA & fi           & $0.48\pm0.03$ & K5   & $9.229\pm0.030$ & 19 02 22.1 & -36 55 41 & $-0.1\pm0.3$ & 0.5 & 3 \\
Kn H$\alpha$14$^{7}$ & CrA & $-27\pm13$ & $0.53\pm0.08$ & M0   & $8.446\pm0.017$ & 19 02 33.1 & -36 58 21 & $10.5\pm4.4$ & 12 & 9 \\
\hline\hline
\end{tabular}
\label{tab:singleIII} 
$^1$ em=emission, abs=absorption fi=filled in \\
$^2$ V2503 Oph (Haro 1-4), V2129 Oph (SR 9), HBC 268 (Haro 1-16), S CrA
     also observed by Melo et al. (\cite{melo03}) and found to be single \\ 
$^3$ V2503 Oph (Haro 1-4) : binary with separation of $0 \farcs 72$  (Ghez et al. \cite{ghez93}) \\
$^4$ V2129 Oph (SR 9) : binary with separation of $0 \farcs 59$ (Ghez et al. \cite{ghez93}) \\
$^5$ HD173148 : binary with separation of $1 \farcs 1$  \\
$^6$ S CrA : binary with separation of $1 \farcs 41\pm0 \farcs 06$  (Ghez et al. \cite{ghez97}) \\
$^7$ Kn H$\alpha$14 : this might be a spectroscopic binary \\
\end{table*}


   \section{Discussion and conclusions}

The goal of the survey was to compile a list of young long-period binary
stars suitable for calibrating the evolutionary tracks by measuring the
masses by combining spectroscopic and VLTI data. Although it was
not our intention to carry out a statistical compleate survey, the
survey does contains some useful informtion on the frequency of young
binaries.  Mathieu (\cite{mathieu94}) listed in his classical review
article only 3 pms-spectroscopic binaries located in the nearby
southern star-forming regions (Chamaeleon, Corona Australis, Lupus,
Sco-Cen, $\rho$ Ophiuchi) with orbital periods longer than 50 days, and
4 short period systems in these regions. Another short-period binary
system (\object{RXJ1108.8-7519}) was found by Covino et
al. (\cite{covino97}).  The long period binaries are NTTS162819-2423S
(period 89.1 days, $\rho$ Oph), Wa Oph~1 (=NTTS160814-1857, period 144.7
days, SC), and Haro 1-14c (period 591 days, $\rho$ Oph) (see also
Reipurth et al. \cite{reipurth02}). NTTS162819-2423S and Haro 1-14c
are both members of hierarchical systems. The distance between
the spectroscopic binary Haro 1-14c and Haro 1-14 is $12 \farcs 9$, and
NTTS162819-2423S is separated from spectroscopic binary NTTS162819-2423N
by $6\farcs$.  Additionally to these stars, Melo (\cite{melo03})
identified VW Cha (ChaI), BF Cha (ChaII), CHX 18N (ChaI), and AS 205
(SC) as spectroscopic binaries but did not derive the orbital periods of
these objects. However, we regard VW Cha not as a spectroscopic
binary. Another spectroscopic binary found is \object{RXJ1603.8-3938}
(Guenther et al. \cite{guenther01}).  Thus, in all previous surveys, 12
SBs were found, 6 have periods longer than 50 days, of two are
triple stars.  To these we add 7 long-period SBs, and 3 short-period
ones.  These short-period systems will be discussed in a forthcoming
paper (Covino et al. \cite{covino07}).  One of the newly found
long-period systems is a triple star. The newly found tripple
system is hirarchical. It consits of two components which are separated
by $0 \farcs 26$ $\pm 0\farcs 03$, were one of these components is a
binary with an orbital period of possibly 16 days. If we count the
triple star as two binaries, because it consist of a long and a short
period system, the total number of SBs in these regions is 25.  By
adding in the previously known binaries, the sample comprises 120 stars
in total.  Thus, the frequency of SBs is $20\pm5\%$. This number can now
be compared with the results by Duquennoy and Mayor
(\cite{duquennoy91}), who found a frequency of binaries with periods
less than 3000 days amongst old G-dwarfs in the solar neighbourhood of
$21\pm4\%$. The frequency of old and young binaries thus is roughly the
same.

%


\begin{acknowledgements}

We are grateful to the user support group of ESO/La Silla. EC and JMA
acknowledge financial support from INAF and Italian MIUR. This research
has made use of the SIMBAD database, operated at CDS, Strasbourg,
France.  This publication makes use of data products from the Two Micron
All Sky Survey, which is a joint project of the University of
Massachusetts and the Infrared Processing and Analysis Center/California
Institute of Technology, funded by the National Aeronautics and Space
Administration and the National Science Foundation. The authors would
also like to thank Andrea Mehner for critically reading the text, and
the anonymous referee for improoving it.

\end{acknowledgements}


\begin{thebibliography}{}

\bibitem[2003]{alencar03} 
         Alencar, S.H.P., Melo, C.H.F., Dullemond, C.P., Andersen, J., Batalha, C.,
         Vaz, L.P.R., \& Mathieu, R.D. 2003, \aap, 409, 1037 

\bibitem[1998]{baraffe98}
         Baraffe, I., Chabrier, G., Allard, F., \& Hauschildt, P.H. 1998, \aap, 337, 403

\bibitem[1999]{bertout99}
         Bertout, C., Robichon, N., \& Arenou, F. 1999, \aap, 352, 574

\bibitem[1994]{beuzit94}
        Beuzit, J.L., Hubin, N., Gendron, E., Demailly, L., Gigan, P.,
        Lacombe, F., Chazallet, F., Rabaud, D., \& Rousset, G. 1994, in:
        {\it Adaptive Optics in Astronomy}, eds. M.A. Ealey \&
        F. Merkle, Proc. SPIE, 2201 955

\bibitem[2005]{boden05}
         Boden, A.F., Sargent, A.I., Akeson, R.L., Carpenter, J.M., Torres, G., Latham, D.W., 
         Soderblom, D.R., Nelan, E., 
         Franz, O.G., \& Wasserman, L.H. 2005, \apj, 635, 442 


\bibitem[1978]{brault78}
         Brault, J.W. 1978, in: Proceedings of JOSO Workshop, G. Godoli,
         G. Noci, \& A. Righini, Arcetri: Osser. Mem. Osserv. Astrofis., 
         Number 106, 33

\bibitem[1996]{brandner96}
         Brandner, W., Alcal\'a, J.M., Kunkel, M., Moneti, A., \& Zinnecker, H.
         1996, \aap, 307, 121

\bibitem[1997]{burrows97}
         Burrows, A., et al. 1997, \apj, 491, 856

\bibitem[1998]{casey98}
         Casey, B.W., Mathieu, R.D., Vaz, L.P.R., Andersen, J.,
         \& Suntzeff, N.B. 1998, \aj, 115, 1617

\bibitem[2000]{chabrier00}
         Chabrier, G., \& Baraffe, I. 2000, \araa, 38, 337

\bibitem[1981]{chini81}
         Chini, G. 1981, \aap, 99, 346

\bibitem[1997]{covino97}
         Covino, E., Alcala, J. M., Allain, S., Bouvier, J., 
         Terranegra, L., \& Krautter, J. 1997, \aap, 328, 187

\bibitem[2004]{covino04}
         Covino, E., Frasca, A., Alcal{\' a}, J.M., Paladino, R.,
         \& Sterzik, M.F.\ 2004, \aap, 427, 637

\bibitem[2007]{covino07}
         Covino, E. et al. in preparation

\bibitem[2003]{cutri03} Cutri, R.~M., et al.\ 2003, 
         VizieR Online Data Catalog: II/246. Originally published in: 
         University of Massachusetts and Infrared Processing and Analysis Center, 
         (IPAC/California Institute of Technology) (2003)

\bibitem[2000]{dantona00}
         D'Antona, F., Ventura, P., \& Mazzitelli, I. 2000, \apjl, 543, L77

\bibitem[1994]{dantona94} 
         D'Antona, F., \& Mazzitelli, I. 1994, \apjl S, 90, 467

\bibitem[1999]{dezeeuw99}
         de Zeeuw, P.T., Hoogerwerf, R., de Bruijne, J.H.J., Brown, A.G.A.,
         \& Blaauw, A., 1999, \aj, 117, 354

\bibitem[1991]{duquennoy91}
         Duquennoy, A., \& Mayor, M. 1991, \aap 248, 485

\bibitem[2006]{esposito06} 
         Esposito, M., Covino, E., Alcal{\'a}, J.M., Guenther, E.W.,
         \& Schisano, E., supmitted to MNRAS

\bibitem[1994]{forestini94}
         Forestini, M. 1994, \aap, 285, 473

\bibitem[2002]{franco02}
         Franco, G.A.P.  2002, \mnras, 331, 474

\bibitem[1997]{ghez93} 
         Ghez, A.M., Neugebauer, G., \& Matthews, K., 1993, \aj, 106, 2005

\bibitem[1997]{ghez97} 
         Ghez, A.M., McCarthy, D.W., Patience, J.L., \& Beck, T.L.\ 1997, 
         \apj, 481, 378

\bibitem[2001]{guenther01}
         Guenther, E.W., Torres, G., Batalha, N., Joergens, V.,
         Neuh{\" a}user, R., Vijapurkar, J., \& Mundt, R. 
         2001, \aap, 366, 965

\bibitem[2005]{guenther05}
         Guenther, E.W., Covino, E., Alcal{\'a}, J.M., Esposito, M., 
         \& Mundt, R.\ 2005, \aap, 433, 629 


\bibitem[2006]{hatzes06}
          Hatzes, A. P., Cochran, W. D., Endl, M., Guenther, E. W., Saar, S. H.,
          Walker, G. A. H., Yang, S., Hartmann, M., Esposito, M., Paulson, D. B.,
          D\"ollinger, M. P. 2006, \aap, 457, 335 

\bibitem[2000]{hatzes00}
         Hatzes, Artie P., Cochran, William D., McArthur, Barbara, Baliunas,
         Sallie L., Walker, Gordon A. H., Campbell, Bruce, Irwin, Alan W., Yang,
         Stephenson, K\"urster, Martin, Endl, Michael, Els, Sebastian, Butler,
         R. Paul, Marcy, Geoffrey W. 2000, \apjl, 544, L145 

\bibitem[1993]{henning93} 
         Henning, T., Pfau, W., Zinnecker, H., \&
         Prusti, T.\ 1993, \aap, 276, 129

\bibitem[1992]{hogeveen92}
         Hogeveen, S.J. 1992, Astrophysics and Space Science, 193, no. 1, 1992, pp. 29-46

\bibitem[1993]{hughes93}
         Hughes, J., Hartigan, P., \& Clampitt, L. 1993, \aj, 105, 571

\bibitem[2001]{joergens01}
         Joergens, V., Guenther, E., Neuh{\" a}user, R., Fern{\' a}ndez, M.,
         \& Vijapurkar, J.\ 2001, \aap, 373, 966

\bibitem[1973] {knacke73}
         Knacke, R. F., Strom, K. M., Strom, S. E., Young, E. T., 
         \& Kunkel, W., 1973, \apj, 179, 847

\bibitem[1998]{knude98}
         Knude, J., \& H\o g, E. 1998, \aap, 338, 897

\bibitem[2001]{koehler01}
         K\"ohler, R. 2001, \aj 122, 3325

\bibitem[2005]{koenig05}
         K\"onig, B., Guenther, E.W., Woitas, J., 
         \& Hatzes, A.P.\ 2005, \aap, 435, 215

\bibitem[2005]{lovis05}
         Lovis, C., Mayor, M., Bouchy, F., Pepe, F., Queloz, D., Santos, N. C., Udry, S., 
         Benz, W., Bertaux, J.-L., Mordasini, C., Sivan, J.-P. 2005,
         \aap, 437, 1121 

\bibitem[1981] {marraco81}
         Marraco, H.G., \& Rydgren, A.E. 1981, \aj, 86, 62


\bibitem[1994]{mathieu94}
         Mathieu, R.D.\ 1994, \araa, 32, 465

\bibitem[2003]{melo03}
         Melo, C.H.F. 2003, \aap, 410, 269

\bibitem[2004]{montalban04}
         Montalb\'an J., D'Antona, F., Kupka, F., \& Heiter, U.
         2004, \aap, 416, 1081

\bibitem[1993]{murdoch93}
         Murdoch, K.A., Hearnshaw, J. B., \& Clark, M., 1993, \apj, 413, 349

\bibitem[2000]{nakajima00}
         Nakajima, Y., Tamura, M., Oasa, Y., \& Nakajima, T. 
         2000, \aj, 119, 873

\bibitem[1996]{padgett96}
         Padget, D.L., 1996, \apj, 471, 847

\bibitem[1992]{palla92}
         Palla, F., \& Stahler, S.W. 1992, \apj, 392, 667

\bibitem[1999]{palla99}
         Palla, F., \& Stahler, S.W. 1999, \apj, 525, 772

\bibitem[2002]{prato02}
         Prato, L., Simon, M., Mazeh, T., McLean, I.S., 
         Norman, D., \& Zucker, S.\ 2002, \apj, 569, 863 

\bibitem[2000]{quist00}
         Quist, C.F., \& Lindegren, L., 2000, \aap, 361, 770

\bibitem[2002]{paulson02}
         Paulson, D.~B., Saar, S.H., Cochran, W.D., 
         Hatzes, A.~P.\ 2002, \aj, 124, 572 

\bibitem[1998]{preibisch98}
         Preibisch, T., Guenther, E., Zinnecker, H., Sterzik, M., 
         Frink, S., Roeser, S.\ 1998, \aap, 333, 619 

\bibitem[1999]{preibisch99}
         Preibisch, T., \& Zinnecker, H. 1999, \aj, 117, 2318

\bibitem[1993]{reipurth93}
         Reipurth, B., \& Zinnecker, H. 1993, \aap, 278, 81

\bibitem[2002]{reipurth02}
         Reipurth, B., Lindgren, H., Mayor, M., Mermilliod, J.-C., 
         \& Cramer, N.\ 2002, \aj, 124, 2813

\bibitem[1998]{saar98}
         Saar, S.~H., Butler, R.P., \& Marcy, G.~W. 1998, \apjl, 498, L153

\bibitem[2003]{satori03}
         Sartori, M.J., Lepine, J.R.D., \& Dias, W.S. 2003, \aap, 404, 913

\bibitem[1997]{siess97}
         Siess, L., Forestini, M., \& Dougados, C. 1997, \aap, 324, 556

\bibitem[2006]{silverstone06}
         Silverstone, M.D., Meyer, M.R., Mamajek, E.E., Hines, D.C.,
         Hillenbrand, L.A., Najita, J., Pascucci, I., Bouwman, J.,
         Kim, J.S., Carpenter, J.M., Stauffer, J.R., Backman, D.E.,
         Moro-Martin, A., Henning, Th.,  Wolf, S., Brooke, T.Y.
         \& Padgett, D.L., 2006, \apj 639, (astro-ph/0511250)

\bibitem[2000]{simon00}
         Simon, M., Dutrey, A., \& Guilloteau, S. 2000, \apj, 545, 1034

\bibitem[2004]{simon04}
        Simon, M., \& Prato, L. 2004, \apjl, 613, L69

\bibitem[2006]{skrutie06} 
         Skrutskie, M. F., Cutri, R. M., Stiening, R. et al., 2006, 
         \aj, 131, 1163

\bibitem[2002]{song02}
         Song, I., Bessell, M.S., Zuckerman, B.\ 2002, \aap, 385, 862 

\bibitem[2004]{stassun04}
         Stassun, K.G., Mathieu, R.D., Vaz, L.P.R., Stroud, N.,
         \& Vrba, F.J.\ 2004, \apjs, 151, 357

\bibitem[2001]{steffen01}
         Steffen, A.T.,  Mathieu, R.D., Lattanzi, M.G., Latham, D.W., Mazeh, T., 
         Prato, L., Simon, M., Zinnecker, H., \& Loreggia, D.
         2001, \aj, 122, 997

\bibitem[2007]{stempels07} 
         Stempels, H.~C., Gahm, G.F., \& Petrov, P.P.\ 2007, \aap, 461, 253

\bibitem[1994]{swenson94}
         Swenson, F.J., Faulkner, J., Iglesias, C.A., Rogers, F.J.,
         \& Alexander, D.R. 1994, \apjl, 422, L79

\bibitem[2003]{takami03}
         Takami, M., Bailey, J., \& Chrysostomou, A., 2003,  \aap, 397, 675-691

\bibitem[2000]{teixeira00}
         Teixeira, R., Ducourant, C., Sartori, M.J., Camargo, J.I.B.,
         P\'eri\'e, J.P., L\'epine, J.R.D., \& Benevides-Soares, P.,
         2000, \aap, 361, 1143

\bibitem[2000]{torres00}
         Torres, C.A.O., da Silva, L., Quast, G. R., de la Reza, R., 
         \& Jilinski, E. 2000, \aj, 120, 1410

\bibitem[2003]{torres03}
         Torres, G., Guenther, E.W., Marschall, L.A., 
         Neuh{\"a}user, R., Latham, D.W., \& Stefanik,
         R.P.\ 2003, \aj, 125, 825 

\bibitem[1999]{tout99}
        Tout, C.A., Livio, M., \& Bonnell, I.A. 1999, \mnras, 310, 360

\bibitem[1998]{wallace98}
         Wallace, L., Hinkle, K., \& Livingston, W., 1998, An atlas of the
         spectrum of the solar photosphere from 13,500 to 28,000 $cm^{-1}$
         (3570 to 7405 \AA ), Publisher: Tucson, AZ: National Optical
         Astronomy Observatories.

\bibitem[1999]{webb99}
         Webb, R.A., Zuckerman, B., Platais, I., Patience, J., White, R.J., 
         Schwartz, M.J., \& McCarthy, C. 1999, \apj, 512, L63

\bibitem[2004]{weinberger04}
         Weinberger, A.J., Becklin, E.E., Zuckerman, B., \& Song, I. 
         2004, \aj, 127, 2246

\bibitem[1974]{whittet74}
         Whittet, D.C.B. 1974, \mnras, 168, 371 

\bibitem[1997]{whittet97} 
         Whittet, D.C.B., Prusti, T., Franco, G.A.P., Gerakines, P.A.,
         Kilkenny, D., Larson, K.A., \& Wesselius, P.R. 1997, \aap, 327, 1194

\bibitem[1998]{wichmann98}
         Wichmann, R., Bastian, U., Krautter, J., Jankovics, I., \& Ruciski,
         S.M. 1998, \mnras, 301, L39

\bibitem[2003]{wuchterl03}
         Wuchterl, G., Tscharnuter, W.M. 2003, \aap, 398, 1081

\end{thebibliography}
\end{document}